\documentclass[aps,prd,twocolumn,twoside,preprintnumbers,showpacs,showkeys,byrevtex,floatfix,nofootinbib,amsmath,amssymb]{revtex4}

\usepackage{amscd,array,dsfont,graphicx,mathtools,multirow}
\usepackage[bookmarksnumbered,linktocpage,pdfstartview=FitH]{hyperref}
\usepackage[all]{hypcap}

\newcolumntype{M}[1]{>{$}{#1}<{$}}

\DeclarePairedDelimiter{\abs}{\lvert}{\rvert}

\DeclareMathOperator{\Tr}{Tr}
\DeclareMathOperator{\sgn}{sgn}
\DeclareMathOperator{\diag}{diag}
\DeclareMathOperator{\Det}{Det}
\DeclareMathOperator{\Aut}{Aut}
\DeclareMathOperator{\Str}{Str}

\newcommand{\half}{\tfrac{1}{2}}
\newcommand{\rep}[1]{\mathbf{#1}}
\newcommand{\cP}{\mathcal{P}}
\newcommand{\cQ}{\mathcal{Q}}
\newcommand{\cJ}{\mathcal{J}}

\begin{document}

\title{Black holes admitting a Freudenthal dual}
\author{L. Borsten}
\email[]{leron.borsten@imperial.ac.uk}
\author{D. Dahanayake}
\email[]{duminda.dahanayake@imperial.ac.uk}
\author{M. J. Duff}
\email[]{m.duff@imperial.ac.uk}
\author{W. Rubens}
\email[]{william.rubens06@imperial.ac.uk}
\affiliation{The Blackett Laboratory, Imperial College London, Prince Consort Road, London SW7 2BZ, U.K.}
\date{\today}

\begin{abstract}

The quantised charges $x$ of four dimensional stringy black holes may be assigned to elements of an integral Freudenthal triple system whose automorphism group is the corresponding U-duality and whose U-invariant quartic norm $\Delta(x)$ determines the lowest order entropy. Here we introduce a \emph{Freudenthal duality} $x \to \tilde{x}$, for which $\tilde{\tilde{x}}=-x$. Although distinct from U-duality it  nevertheless leaves $\Delta(x)$ invariant. However, the requirement that $\tilde x$ be integer restricts us to the subset of black holes for which $\Delta(x)$ is necessarily a perfect square. The issue of higher-order corrections remains open as some, but not all, of the discrete U-duality invariants are Freudenthal invariant. Similarly, the quantised charges $A$ of five dimensional black holes and strings may be assigned to elements of an integral Jordan algebra, whose cubic norm $N(A)$ determines the lowest order entropy. We introduce an analogous \emph{Jordan dual} $A^\star$, with $N(A)$ necessarily a perfect cube, for which $A^{\star\star}=A$ and which leaves $N(A)$ invariant. The two dualities are related by a 4D/5D lift.

\end{abstract}



\preprint{Imperial/TP/2009/mjd/1}

\maketitle

\section{Introduction}
\label{sec:Introduction}

The purpose of this paper is to introduce two new dualities, distinct from U-duality, which act on black hole charges in 4D and 5D and which leave the lowest order entropy invariant. Some, but not all, of the other discrete U-duality invariants are also conserved, so the question of higher order corrections remains open.

It is well known that the four dimensional supergravities that arise from string and M-theory, such as the $\mathcal{N}=2$ $STU$, $\mathcal{N}=2$ ``magic'', $\mathcal{N}=4$ heterotic and $\mathcal{N}=8$ M/Type II, may all be described by a Freudenthal triple system (FTS) $\mathfrak{M(J)}$ \cite{Freudenthal:1954,Brown:1969,Gunaydin:1983rk, Ferrara:1997ci, Ferrara:1997uz,Gunaydin:2005gd,Bellucci:2006xz,Ferrara:2006yb,Pioline:2006ni,Borsten:2008wd}, where $\mathfrak{J}$ is a cubic Jordan algebra underlying the corresponding 5D supergravity \cite{Gunaydin:1983rk,Gunaydin:1983bi,Gunaydin:1984ak}. The corresponding continuous U-duality is given by the automorphism group $\Aut(\mathfrak{M(J)})$, e.g.\ $E_{7(7)}$ in the case of $\mathcal{N}=8$ \cite{Cremmer:1979up}. The FTS admits a skew-symmetric  bilinear form $\{x,y\}$, a quartic form $\Delta(x,y,z,w)$ and a trilinear operator $T(x,y,z)$, defined by $\{T(x,y,z),w\}=2\Delta(x,y,z,w)$. To lowest order, the extremal non-rotating black hole entropy is given by
\begin{equation}
S_4=\pi\sqrt{|\Delta(x)|},
\end{equation}
where $\Delta(x)= \Delta(x,x,x,x)$.  ``Large'' BPS, ``small'' BPS and large non-BPS correspond to $\Delta(x) >0$, $\Delta(x) =0$ and $\Delta(x) <0$, respectively.  In this continuous case, the black hole entropy, U-duality orbits and generating solutions are well understood \cite{Cvetic:1995bj, Cvetic:1995uj,Cvetic:1995kv,Kallosh:1996uy,Cvetic:1996zq,Andrianopoli:1997pg,Ferrara:1997uz,Lu:1997bg,Bertolini:1999je,Bertolini:2000ei,Bertolini:2000yaa,Bellucci:2006xz,Andrianopoli:2006ub,Borsten:2008wd}.

In the fully quantised string theory, however, black hole charges $x$ must be integer valued and hence assigned to elements of an \emph{integral} FTS $\mathfrak{M(J)}$ where $\mathfrak{J}$ is an \emph{integral} cubic Jordan algebra \cite{Elkies:1996,Gross:1996,Krutelevich:2002, Krutelevich:2004}.    The corresponding U-duality is given by the discrete  automorphism group $\Aut(\mathfrak{M(J)})$, e.g.\ $E_{7(7)}( \mathds{Z})$ in the case of $\mathcal{N}=8$ \cite{Hull:1994ys}, with $x$ transforming as a $\rep{56}$. In particular, $\Delta(x)$ is now quantised:
\begin{equation}
\Delta(x) \in \{0,1\} \mod 4.
\end{equation}
From a mathematical point of view, much less is known about the integral case. For example, we shall see that the general classification of U-duality orbits in $D=4$ is lacking, except for the special class of \emph{projective} black holes. The class of projective FTS elements is of particular relevance to recent developments in number theory \cite{Bhargava:2004, Krutelevich:2004}.

Here we introduce the  \emph{Freudenthal dual} or F-dual, defined for large BPS and non-BPS black holes by
\begin{equation}\label{eq:F-dual}
\tilde{x}=T(x) |\Delta(x)|^{-1/2},
\end{equation}
where $T(x)= T(x,x,x) \in \mathfrak{M(J)}$. Requiring that $\tilde x$ is integer therefore restricts us to that subset of black holes for which $|\Delta(x)|$ is a perfect square and for which $|\Delta(x)|^{1/2}$ divides $T(x)$:
\begin{equation}\label{eq:d4}
d_4(x)=\left [\frac{d_3(x)}{d_1(\tilde x)}\right]^2,
\end{equation}
where $d_1(x)=\gcd (x)$, $d_3(x)= \gcd (T(x))$ and $d_4(x)=|\Delta(x)|$.
Applying the F-duality once more yields
\begin{equation}\label{double-dual}
\tilde{\tilde{x}}=-x.
\end{equation}
Despite the non-polynomial nature of the transformation  \eqref{eq:F-dual}, the F-dual scales linearly in the sense that
\begin{equation}
{\tilde{x}(nx)}= n {\tilde x}(x), \quad n\in \mathds{Z}.
\end{equation}

The U-duality integral invariants $\{x,y\}$ and $\Delta(x,y,z,w)$ are not generally invariant under F-duality but $ \{\tilde{x},x\}$, $\Delta(x)$,  and hence the lowest-order black hole entropy, are invariant. However, higher order corrections may also depend on discrete U-duality invariants involving the various $\gcd$s \cite{Sen:2007qy,Banerjee:2007sr,Banerjee:2008ri,Sen:2008ta,Sen:2008sp}.  Under F-duality certain discrete U-duality invariants are conserved while others are not necessarily, as is discussed in \autoref{sec:Fdualdiscrete}. For example, the product  $d_1(x)d_3(x)$ is invariant but $d_1(x)$ and $d_3(x)$ separately need not be. A 4D black hole is called \emph{primitive} if $d_1(x)=1$, so the F-dual of a primitive black hole need not itself be primitive.

As described in \autoref{sec:J}, the FTS divides black holes into five distinct ranks or orbits. Though F-duality \eqref{eq:F-dual} was defined for rank 4 black holes for which both $T$ and $\Delta$ are nonzero, in \autoref{sec:small} we consider extending  to ranks 0, 1 and 2 for which both $T$ and $\Delta$ vanish (but not rank 3 for which $\Delta$ vanishes but not $T$).  However, the apparent lack of uniqueness favours continuing to restrict F-duality to large black holes.

Similar remarks apply to the quantised charges  $A$ of five dimensional black strings and the quantised charges  $B$ of five dimensional  black holes which may be assigned to elements of an integral cubic Jordan algebra $\mathfrak{J}$, whose reduced structure group $\Str_0(\mathfrak{J})$ is the corresponding U-duality, e.g.\ $E_{6(6)}(\mathds{Z})$ in the case of $\mathcal{N}=8$ with $A$ transforming as a $\rep{27}$ and $B$ as a $\rep{27'}$. The Jordan algebra admits a trace bilinear form $\Tr(X,Y)$, a cubic norm $N(X,Y,Z)$ and a quadratic adjoint map $X^\sharp$ uniquely defined by $\Tr(X^\sharp, Y) = 3N(X, X, Y)$.  To lowest order, the extremal non-rotating black hole and black string entropies are given respectively by
\begin{equation}\label{eq:5-entropy}
\begin{split}
S_{5(\text{black string})}&=2\pi\sqrt{|N(A)|},\\
S_{5(\text{black hole})}  &=2\pi\sqrt{|N(B)|},
\end{split}
\end{equation}
where $N(A)= N(A,A,A) \in \mathds{Z}$. Large BPS and small BPS correspond to $N \neq 0$, and $N=0$, respectively.

Here we also introduce the \emph{Jordan dual} or J-dual, defined for ``large'' black strings and holes by
\begin{equation}\label{eq:d5dualx}
{A}^\star=A^\sharp N(A)^{-1/3},\qquad  {B}^\star= B^\sharp N(B)^{-1/3},
\end{equation}
where we take the real root as implied by the notation. Requiring that ${A}^\star$ and ${B}^\star$ are integers therefore restricts us to that subset of black holes for which $N(A)$ and $N(B)$ are perfect cubes and for which $N(A)^{1/3}$ divides $A^\sharp$ and $N(B)^{1/3}$ divides $B^\sharp$
\begin{align}
d_3({A})&=\left[ \frac{d_2(A)}{d_1({A}^\star)}\right]^3,&
d_3({B})&=\left[ \frac{d_2(B)}{d_1({B}^\star)}\right]^3,
\end{align}
where $d_1(A)=\gcd (A)$, $d_2(A)= \gcd (A^\sharp)$, $d_3(A)=N(A)$ and similarly for $B$.  Applying the J-duality once more yields
\begin{align}
{A}^{\star\star}&=A, & {B}^{\star\star}&=B.
\end{align}
Despite the non-polynomial nature of the transformation \eqref{eq:d5dualx}, the J-dual scales linearly in the sense that
\begin{equation}
\begin{split}
{A^\star(nA)}&= n {A^\star}(A), \\
{B^\star(nB)}&= n {B^\star}(B), \quad n\in \mathds{Z}.
\end{split}
\end{equation}

The U-duality integral invariants $\Tr(X,Y)$ and $N(X,Y,Z)$ are not generally invariant under Jordan duality but $\Tr(X^\star, X)$,  $N(X)$ and hence the lowest-order black hole and black string entropy, are invariant.  However, higher order corrections may also depend on discrete U-duality invariants involving the various $\gcd$s \cite{Sen:2007qy,Banerjee:2007sr,Banerjee:2008ri,Sen:2008ta,Sen:2008sp}. Under J-duality certain discrete U-duality invariants are conserved while others are not necessarily, as is discussed in \autoref{sec:Jaction}. For example, the product  $d_1(A)d_2(A)$ is invariant but $d_1(A)$ and $d_2(A)$ separately need not be. A 5D black hole/string is called \emph{primitive} if $d_1=1$, so the J-dual of a primitive black hole/string need not itself be primitive.

As described in \autoref{sec:J}, the Jordan algebra divides black strings/holes into four distinct ranks or orbits. Though J-duality \eqref{eq:d5dualx} was defined for rank 3, for which both $A^\sharp$ and $N(A)$ are nonzero, in \autoref{sec:Smith} we consider extending the definition to ranks 0 and 1, for which both $A^\sharp$ and $N(A)$ vanish, (but not rank 2 for which $N(A)$ vanishes but not $A^\sharp$). However, the apparent lack of uniqueness favours continuing to restrict J-duality to large black holes/strings.

Many of our results simplify if we confine our attention to the NS-NS sector, which is interesting in its own right for the heterotic and $STU$ black holes. This is treated in \autoref{sec:NS}, where \emph{inter alia} we answer yes to the question posed in \cite{Sen:2008sp}: Is a general  $D=4$, $\mathcal{N}=8$ black hole always U-duality related to one with only NS-NS charges?

The 4D/5D lift \cite{Gaiotto:2005gf} associates a rotating 5D black hole to a non-rotating 4D black hole. In \autoref{sec:FJlift} we show that two black holes related by F-duality in 4D are related by J-duality when lifted to 5D.

In \autoref{sec:conclusions} we examine the all-important question of the invariance of the exact entropies under F and J dualities.  In the special 5D and \emph{projective} 4D cases where all U-duality invariants are preserved, the exact entropy is F and J dual invariant, but in a trivial way: the transformations can always be undone by a U-duality. In the 4D non-projective case, the question of U-equivalence remains open because of the inability to ``reverse engineer'' the black holes charges given their (known) U-duality invariants.  In the 4D and 5D cases where not all U-duality invariants are preserved, there is insufficient information and further research is required.

\section{Review of Jordan algebras and the Freudenthal triple system}
\label{sec:FTS}

\subsection{Jordan algebras and 5D black holes}
\label{sec:J}

A Jordan algebra $\mathfrak{J}$ \cite{Jordan:1933a,Jordan:1933b,Jordan:1933vh,Jacobson:1961,Jacobson:1968} is vector space defined over a ground field $\mathds{F}$ equipped with a bilinear product satisfying,
\begin{equation}\label{eq:Jid}
\begin{split}
X\circ Y           &=Y\circ X, \\
X^2\circ (X\circ Y)&=X\circ (X^2\circ Y), \quad\forall\ X, Y \in \mathfrak{J}.
\end{split}
\end{equation}
For our purposes the relevant Jordan algebras are all examples of the class of \emph{integral cubic} Jordan algebras \cite{Elkies:1996,Gross:1996, Krutelevich:2002,Krutelevich:2004}. An integral cubic Jordan algebra comes equipped with a cubic form $N:\mathfrak{J}\to\mathds{Z}$, satisfying $N(\lambda X)=\lambda^3N(X), \quad \forall\ \lambda \in \mathds{Z},\ X\in \mathfrak{J}$. Additionally, there is an element $c\in\mathfrak{J}$ satisfying $N(c)=1$, referred to as a \emph{base point}. There is a general prescription for constructing cubic Jordan algebras, due to Freudenthal, Springer and Tits \cite{Springer:1962, McCrimmon:1969, McCrimmon:2004}, for which all the properties of the Jordan algebra are essentially determined by the cubic form.  We sketch this construction here, following closely the conventions of \cite{Krutelevich:2004, McCrimmon:2004}.

Let $V$ be a vector space equipped with both a cubic norm, $N:V\to \mathds{Z}$, satisfying $N(\lambda X)=\lambda^3N(X), \ \forall\ \lambda \in \mathds{Z},\ X\in V$, and a base point $c\in V$ such that $N(c)=1$. If $N(X, Y, Z)$, referred to as the full \emph{linearisation} of $N$, defined by
\begin{equation}
\begin{gathered}
N(X, Y, Z):=\\
\begin{split}
\tfrac{1}{6}[&\phantom{-\;\,}N(X+ Y+ Z)\\&-N(X+Y)-N(X+ Z)-N(Y+ Z)\\&+N(X)+N(Y)+N(Z)]
\end{split}
\end{gathered}
\end{equation}
is trilinear then  one may define the following four maps,
\begin{subequations}\label{eq:cubicdefs}
\begin{enumerate}
\item The trace,
    \begin{equation}
    \Tr(X)=3N(c, c, X),
    \end{equation}
\item A quadratic map,
    \begin{equation}
    S(X)=3N(X, X, c),
    \end{equation}
\item A bilinear map,
    \begin{equation}
    S(X, Y)=6N(X, Y, c),
    \end{equation}
\item A trace bilinear form,
    \begin{equation}\label{eq:tracebilinearform}
    \Tr(X, Y)=\Tr(X)\Tr(Y)-S(X, Y).
    \end{equation}
\end{enumerate}
\end{subequations}
A cubic Jordan algebra $\mathfrak{J}$ with multiplicative identity $\mathds{1}=c$ may be derived from any such vector space if $N$ is \emph{Jordan cubic}, that is:
\begin{enumerate}
\item The trace bilinear form \eqref{eq:tracebilinearform} is non-degenerate.
\item The quadratic adjoint map, $\sharp\colon\mathfrak{J}\to\mathfrak{J}$, uniquely defined by $\Tr(X^\sharp, Y) = 3N(X, X, Y)$, satisfies
    \begin{equation}\label{eq:Jcubic}
    (X^{\sharp})^\sharp=N(X)X, \quad \forall X\in \mathfrak{J}.
    \end{equation}
\end{enumerate}
The Jordan product is then defined using,
\begin{equation}
X\circ Y = \half\big(X\times Y+\Tr(X)Y+\Tr(Y)X-S(X, Y)\mathds{1}\big),
\end{equation}
where, $X\times Y$ is the linearisation of the quadratic adjoint,
\begin{equation}\label{eq:FreuProduct}
X\times Y = (X+Y)^\sharp-X^\sharp-Y^\sharp.
\end{equation}
Finally, the Jordan triple product is defined as
\begin{equation}\label{eq:Jtripleproduct}
\{X,Y,Z\}=(X\circ Y)\circ Z + X\circ (Y\circ Z)-(X\circ Z)\circ Y.
\end{equation}
While in general an integral Jordan algebra is not closed under the Jordan product,  the cubic norm and trace bilinear form are integer valued, which are the crucial properties for our purposes. Moreover,  $\mathfrak{J}$  is closed under the quadratic adjoint map and its linearisation as required.

Important examples include the sets of $3\times 3$ Hermitian matrices, which we denote as $J_{3}^{\mathds{A}}$,  defined over the four division algebras $\mathds{A=R,C,H}$ or $\mathds{O}$  (or their split signature cousins) with Jordan product $X\circ Y=\half(XY+YX)$, where $XY$ is just the conventional matrix product. See \cite{Jacobson:1968} for a comprehensive account. In addition  there is the infinite sequence of \emph{spin factors} $\mathds{Z}\oplus Q_n$, where $Q_n$ is an $n$-dimensional vector space over $\mathds{Z}$ \cite{Jacobson:1961,Jacobson:1968,Krutelevich:2004,McCrimmon:1969,Baez:2001dm}.

The \emph{structure} group, $\Str(\mathfrak{J})$, is composed of all linear bijections on $\mathfrak{J}$ that leave the cubic norm $N$ invariant up to a fixed scalar factor,
\begin{equation}
N(g(X))=\lambda N(X), \quad \forall\ g\in \Str(\mathfrak{J}).
\end{equation}
The \emph{reduced structure} group $\Str_0(\mathfrak{J})$ leaves the cubic norm invariant and therefore consists of those elements in $\Str(\mathfrak{J})$ for which $\lambda =1$ \cite{Schafer:1966, Jacobson:1968, Brown:1969}. The usual concept of matrix rank may be generalised to cubic Jordan algebras and is invariant under both $\Str(\mathfrak{J})$ and $\Str_0(\mathfrak{J})$ \cite{Jacobson:1961, Krutelevich:2004}. See \autoref{tab:jordanrank}.
\begin{table}
\caption[Partition of $\mathfrak{J}$ into four orbits of $\Str_0(\mathfrak{J})$ or ranks]{Partition of the space $\mathfrak{J}$ into four orbits of $\Str_0(\mathfrak{J})$ or ranks.\label{tab:jordanrank}}
\begin{ruledtabular}
\begin{tabular}{ccc*{3}{M{c}}ccc}
& \multirow{2}{*}{Rank} & & \multicolumn{3}{c}{Condition} & & \multirow{2}{*}{$\mathcal{N}=8$ BPS} & \\
\cline{3-7}
&                       & & A     & A^\sharp & N(A)       & &                                      & \\
\hline
& 0                     & & =0    & =0       & =0         & & 1                                    & \\
& 1                     & & \neq0 & =0       & =0         & & 1/2                                  & \\
& 2                     & & \neq0 & \neq0\   & =0         & & 1/4                                  & \\
& 3                     & & \neq0 & \neq0\   & \neq0      & & 1/8                                  &
\end{tabular}
\end{ruledtabular}
\end{table}
\begin{table*}
\caption[Partition of $\mathfrak{M(J)}$ into five orbits of $\Aut(\mathfrak{M(J)})$ or ranks]{Partition of the space $\mathfrak{M(J)}$ into five orbits of $\Aut(\mathfrak{M(J)})$ or ranks.}\label{tab:FTSrank}
\begin{ruledtabular}
\begin{tabular}{ccc*{4}{M{c}}ccc}
& \multirow{2}{*}{Rank} & & \multicolumn{4}{c}{Condition}                     & & \multirow{2}{*}{$\mathcal{N}=8$ BPS} & \\
\cline{3-8}
&                       & & x     & 3T(x,x,y)+\{x,y\}x & T(x,x,x) & \Delta(x) & &                                      & \\
\hline
& 0                     & & =0    & =0\ \forall y      & =0       & =0        & & 1                                    & \\
& 1                     & & \neq0 & =0\ \forall y      & =0       & =0        & & 1/2                                  & \\
& 2                     & & \neq0 & \neq0              & =0       & =0        & & 1/4                                  & \\
& 3                     & & \neq0 & \neq0              & \neq0    & =0        & & 1/8                                  & \\
& 4                     & & \neq0 & \neq0              & \neq0    & >0        & & 1/8                                  & \\
& 4                     & & \neq0 & \neq0              & \neq0    & <0        & & 0                                    &
\end{tabular}
\end{ruledtabular}
\end{table*}

To lowest order, the extremal non-rotating black hole and black string entropies are given respectively by
\begin{equation}
\begin{split}
S_{5(\text{black string})}&=2\pi\sqrt{N(A)}, \\
S_{5(\text{black hole})}  &=2\pi\sqrt{N(B)}.
\end{split}
\end{equation}
Large BPS  black holes and strings correspond to rank 3 with $N(A), N(B) \neq 0$ and small BPS correspond to ranks 1 and 2 with $N(A), N(B)=0$. In \autoref{tab:FTSrank} we have listed the fraction of unbroken supersymmetry for the $\mathcal{N}=8$ case.

The Dirac-Schwinger quantisation condition for an electric black hole and a magnetic  string  with charges $A,B$  in the Jordan language is given by
\begin{equation}\label{eq:DSJ}
\Tr(A, B)\in \mathds{Z}.
\end{equation}

\subsection{The Freudenthal triple system and 4D black holes}
\label{sec:F}

Given an integral cubic Jordan algebra $\mathfrak{J}$, one is able to construct an integral FTS by defining the vector space $\mathfrak{M(J)}$,
\begin{equation}
\mathfrak{M(J)}=\mathds{Z\oplus Z}\oplus \mathfrak{J\oplus J}.
\end{equation}
An arbitrary element $x\in \mathfrak{M(J)}$ may be written as a ``$2\times 2$ matrix'',
\begin{equation}
x=\begin{pmatrix}\alpha&A\\B&\beta\end{pmatrix},\text{\ where\ }\alpha,\beta\in\mathds{Z}\text{\ and\ }A,B\in\mathfrak{J}.
\end{equation}
For convenience we identify the quantity
\begin{equation}
\kappa(x):=\half(\alpha\beta-\Tr(A,B)).
\end{equation}
The FTS comes equipped with a non-degenerate bilinear antisymmetric quadratic form, a quartic form and a trilinear triple product \cite{Freudenthal:1954,Brown:1969,Faulkner:1971, Ferrar:1972, Krutelevich:2004}:
\begin{subequations}
\begin{enumerate}
\item Quadratic form $ \{x, y\}$: $\mathfrak{M(J)}\times \mathfrak{M(J)}\to \mathds{Z}$
    \begin{equation}\label{eq:bilinearform}
    \begin{gathered}
    \{x, y\}=\alpha\delta-\beta\gamma+\Tr(A,D)-\Tr(B,C),\\
    \text{where\ }x=\begin{pmatrix}\alpha&A\\B&\beta\end{pmatrix},\  y=\begin{pmatrix}\gamma&C\\D&\delta\end{pmatrix}.
    \end{gathered}
    \end{equation}
\item Quartic form $q:\mathfrak{M(J)}\to \mathds{Z}$
    \begin{equation}\label{eq:quarticnorm}
    \begin{split}
    \Delta (x)=-4\big[&\alpha N(A)+\beta N(B)\\
                      &\kappa(x)^2-\Tr(A^\sharp, B^\sharp)\big].
    \end{split}
    \end{equation}
    The quartic norm $\Delta(x)$ is either $4k$ or $4k+1$ for some $k\in\mathds{Z}$.
\item Triple product $T:\mathfrak{M(J)}\times
    \mathfrak{M(J)}\times\mathfrak{M(J)}\to\mathfrak{M(J)}$ which is uniquely defined by
    \begin{equation}
    \{T(x, y, w), z\}=2\Delta(x, y, w, z),
    \end{equation}
    where $\Delta(x, y, w, z)$ is the full linearisation of $\Delta(x)$ such that $\Delta(x, x, x, x)=\Delta(x)$. For future convenience we present here an explicit form for $T(x)=T(x,x,x)$:
    \begin{widetext}
    \begin{equation}\label{eq:Tofx}
    T(x)= \begin{pmatrix}T_\alpha&T_A\\T_B&T_\beta\end{pmatrix}= 2\begin{pmatrix}-\alpha\kappa(x)-N(B)&-(\beta B^\sharp-B\times A^\sharp)+\kappa(x)A\\
    (\alpha A^\sharp-A\times B^\sharp)-\kappa(x)B&\beta\kappa(x)+N(A)\end{pmatrix}.
    \end{equation}
    \end{widetext}
\end{enumerate}
\end{subequations}
Note that all the necessary definitions, such as the cubic and trace bilinear forms, are inherited from the underlying Jordan algebra $\mathfrak{J}$.

Of particular importance to our discussion is the \emph{automorphism group} $\Aut(\mathfrak{M(J)})$, which is given by the set of all invertible $\mathds{Z}$-linear transformations which leave both $\{x, y\}$ and $\Delta (x, y, w, z)$ invariant \cite{Brown:1969}. Note, for any transformation $\sigma\in \Aut(\mathfrak{M(J)})$ we have
\begin{equation}
T(\sigma(x), \sigma(y), \sigma(w))=\sigma(T(x, y, w)).
\end{equation}
$\Aut(\mathfrak{M(J)})$ is the U-duality group e.g.\ $E_{7(7)}$ in the case of $\mathcal{N}=8$. The discrete 4D U-duality group is generated by the following three maps \cite{Brown:1969, Krutelevich:2004}:
\begin{widetext}
\begin{subequations}
\begin{align}\label{eq:FreudenthalConstructionTransformations}
\phi(C) &: \begin{pmatrix}\alpha&A\\B&\beta\end{pmatrix} \mapsto \begin{pmatrix}\alpha+(B,C)+(A, C^{\sharp})+\beta N(C)&A+\beta C\\
                                                                                 B+A\times C+\beta C^\sharp  &\beta
                                                                 \end{pmatrix},\\
\psi(D) &: \begin{pmatrix}\alpha&A\\B&\beta\end{pmatrix} \mapsto \begin{pmatrix}\alpha& A+B\times D+\alpha D^\sharp\\
                                                                                B+\alpha D &\beta+(A,D)+(B, C^{\sharp})+\alpha N(C)
                                                                 \end{pmatrix},\\
T(s) &:    \begin{pmatrix}\alpha&A\\B&\beta\end{pmatrix} \mapsto \begin{pmatrix}\lambda^{-1}\alpha & s(A)\\
                                                                                {s'}^{-1}(B) &\lambda\beta
                                                                 \end{pmatrix}
\end{align}
\end{subequations}
\end{widetext}
where $s\in \Str(\mathfrak{J})$  and $s'$ is its adjoint defined with respect to the trace bilinear form, $\Tr(X, s(Y))=\Tr(s'(X), Y)$.

Following \cite{Krutelevich:2004}, the Freudenthal triple systems,  defined by the various Jordan algebras mentioned here, and their associated automorphism groups are summarised in \autoref{tab:FTSsummary}. This table covers most of the black holes of interest: $\mathcal{N}=2$ $STU$, $\mathcal{N}=2$ coupled to $n$ vector multiplets; magic $\mathcal{N}=2$ and $\mathcal{N}=8$.  The heterotic string with $\mathcal{N}=4$ supersymmetry and  $SL(2,\mathds{Z})\times SO(6,22; \mathds{Z})$ U-duality may also be included by using the Jordan algebra $\mathds{Z}\oplus Q_{5,21}$ \cite{Pioline:2006ni,Gunaydin:2009dq}.

The conventional concept of matrix rank may be generalised to Freudenthal triple systems in a natural and $\Aut(\mathfrak{M(J)})$ invariant manner. The rank of an arbitrary element $x\in\mathfrak{M(J)}$ is uniquely defined using the relations in \autoref{tab:FTSrank} \cite{Ferrar:1972, Krutelevich:2004}.
\begin{table*}
\caption[Jordan algebras, corresponding FTSs,  and their associated symmetry groups]{The automorphism  group $\Aut(\mathfrak{M(J)})$ and the dimension of its representation $\dim\mathfrak{M(J)}$ given by the Freudenthal construction defined over the cubic Jordan algebra $\mathfrak{J}$ with dimension $\dim\mathfrak{J}$ and reduced structure group $\Str_0(\mathfrak{J})$. The quantised $\mathcal{N}=8$ theories in 5 and 4 dimensions have U-duality groups $E_{6(6)}(\mathds{Z})$ and $E_{7(7)}(\mathds{Z})$ respectively.}\label{tab:FTSsummary}
\begin{ruledtabular}
\begin{tabular}{c*{5}{M{c}}c}
& \text{Jordan algebra\ }\mathfrak{J}				&\Str_0(\mathfrak{J}) &\dim\mathfrak{J}& \Aut(\mathfrak{M(J)})&\dim\mathfrak{M(J)}&\\
\hline
& \mathds{Z}                                &		-					& 1   & SL(2, \mathds{Z})                         & 4    & \\
& \mathds{Z}\oplus\mathds{Z}                &   SO(1,1, \mathds{Z}) 						& 2   & SL(2, \mathds{Z})\times SL(2, \mathds{Z})			   & 6    & \\
& \mathds{Z}\oplus\mathds{Z}\oplus\mathds{Z}& 	SO(1,1, \mathds{Z})\times SO(1,1, \mathds{Z})	& 3   & SL(2, \mathds{Z})\times SL(2, \mathds{Z})\times SL(2, \mathds{Z}) & 8    & \\
& \mathds{Z}\oplus Q_n                      & 	SO(n-1, 1,\mathds{Z})\times SO(1,1, \mathds{Z})	& n+1 & SL(2, \mathds{Z})\times SO(2,n, \mathds{Z})           & 2n+4 & \\
& J_{3}^{\mathds{Z}}                        & 	SL(3, \mathds{Z})	& 6   & Sp(6,\mathds{Z})                           & 14   & \\
& J_{3}^{\mathds{C}}                        & 	SL(3, \mathds{C})	& 9   & SU(3,3, \mathds{Z})                           & 20   & \\
& J_{3}^{\mathds{H}}                        & 	SU^\star(6, \mathds{Z})									 & 15  & SO^\star(12, \mathds{Z})                           & 32   & \\
& J_{3}^{\mathds{O}}                      	& 	E_{6(-26)}(\mathds{Z})									 & 27  & E_{7(-25)}(\mathds{Z})                           & 56   & \\
& J_{3}^{\mathds{O}^s}                      & 	E_{6(6)}(\mathds{Z})								 & 27  & E_{7(7)}(\mathds{Z})                     & 56   &
\end{tabular}
\end{ruledtabular}
\end{table*}
The rank of any element is invariant under $\Aut(\mathfrak{M(J)})$ \cite{Krutelevich:2004}.

To lowest order, the extremal non-rotating black hole entropy is given by
\begin{equation}\label{eq:4-entropy}
S_4=\pi\sqrt{|\Delta(x)|}.
\end{equation}
Large BPS and large non-BPS black holes correspond to rank 4 with $\Delta(x) >0$ and $\Delta(x) <0$, respectively. Small BPS black holes correspond to ranks 1, 2 and 3 with $\Delta(x) =0$. In \autoref{tab:FTSrank} we have listed the fraction of unbroken supersymmetry for the $\mathcal{N}=8$ case.

The Dirac-Schwinger quantisation condition relating two black holes with charges $x$ and $x'$ within the FTS language is given by
\begin{equation}
\label{eq:DS}
\{x, x'\}\in \mathds{Z}.
\end{equation}

\subsection{The 4D/5D lift}
\label{sec:lift}

Recent work  \cite{Gaiotto:2005gf} has established a simple correspondence relating the entropy of 4D BPS black holes in type IIA theory compactified on a Calabi-Yau $Y$ to the entropy of spinning 5D BPS black holes in M-theory compactified on $Y\times TN_\beta$, where $TN_\beta$ is a Euclidean 4-dimensional Taub-NUT space with NUT charge  $\beta$. Using this  4D/5D lift the electric black hole charge $\cQ$ and spin $\cJ_\beta$ may be identified with the dyonic charges of the 4D black hole giving a precise relationship between the leading order entropy formulae. This relationship has then been used to count the 4D BPS black hole degeneracies in $\mathcal{N}=8$ string theory \cite{Shih:2005qf} exploiting the known results from the analysis of 5-dimensional black holes \cite{Strominger:1996sh, Maldacena:1999bp,Shih:2005qf, Pioline:2005vi,Sen:2007qy,Sen:2008ta,Sen:2008sp}.

This correspondence between the $D=5$ black hole changes $\cQ$ and $\cJ_\beta$ and the $D=4$ electric/magnetic black hole charges is neatly captured in terms of the FTS \cite{Pioline:2006ni}.

Identifying  the black string magnetic charge $\cP$ and black hole electric charge $\cQ$
\begin{equation}
\begin{split}\label{eq:5DPQ}
\cP=B^\sharp -\alpha A,\\
\cQ=A^\sharp -\beta B,
\end{split}
\end{equation}
and the corresponding angular momenta
\begin{equation}
\begin{split}
\cJ_\alpha&= -\half T_\alpha=\phantom{-}\alpha\kappa(x)+N(B),\\
\cJ_\beta &= -\half T_\beta=-\beta\kappa(x)-N(A),
\end{split}
\end{equation}
we find
\begin{equation}\label{eq:cPcQcJdelta}
\Delta (x)=\frac{4}{\alpha^2}\{N(\cP)-{\cJ_\alpha}^2\}=\frac{4}{\beta^2}\{N(\cQ)-{\cJ_\beta}^2\}.
\end{equation}
Hence
\begin{equation}
S_4=\frac{1}{\alpha}S_{5(\text{black string})}=\frac{1}{\beta}S_{5(\text{black hole})},
\end{equation}
where, allowing for rotation,
\begin{equation}
\begin{split}
S_{5(\text{black string})}&=2\pi\sqrt{\abs{N(\cP)-{\cJ_\alpha}^2}},\\
S_{5(\text{black hole})}  &=2\pi\sqrt{\abs{N(\cQ)-{\cJ_\beta}^2}}.
\end{split}
\end{equation}

To prove \eqref{eq:cPcQcJdelta} from a purely Jordan algebraic perspective we begin by using the identity
\begin{equation}
\Tr(X, X^\sharp)=3N(X)
\end{equation}
to write
\begin{equation}
3N(\alpha A-B^\sharp)= \Tr (\alpha A-B^\sharp, (\alpha A-B^\sharp)^\sharp ).
\end{equation}
Then, using
\begin{equation}
(X+Y)^\sharp  = X \times Y + X^\sharp  + Y^\sharp,
\end{equation}
we have
\begin{equation}
\begin{split}\label{eq:NofAplusB}
   &3N(\alpha A-B^\sharp )\\
=\,&\Tr(\alpha A-B^\sharp , (-\alpha A) \times B^\sharp  + \alpha^2 A^\sharp  + N(B)B )\\
=\,&\Tr(\alpha A-B^\sharp , (-\alpha A) \times B^\sharp ) + 3\alpha^3 N(A) \\
   &-\alpha^2 \Tr(A^\sharp , B^\sharp )+\alpha \Tr(A, B)N(B) -3 N(B)^2.
\end{split}
\end{equation}
Finally, using
\begin{equation}
\Tr(X, Y\times Z)=6N(X,Y,Z),
\end{equation}
which may be derived from the definition of the quadratic adjoint
\begin{equation}
\Tr(X^\sharp , Y) = 3N(X, X, Y),
\end{equation}
we see that
\begin{equation}
\begin{split}
 &\Tr (\alpha A-B^\sharp , (-\alpha A) \times B^\sharp )\\
=&\ 6N(\alpha A-B^\sharp ,-\alpha A, B^\sharp )\\
=&\ 2[N(\alpha A-B^\sharp )+N(B^\sharp )-N(\alpha A)].
\end{split}
\end{equation}
Hence, on substituting back into \eqref{eq:NofAplusB} one finds
\begin{equation}\label{eq:NofPid}
\begin{split}
N(\alpha A-B^\sharp )&=\phantom{+}\alpha^3 N(A) -\alpha^2 \Tr(A^\sharp , B^\sharp )\\
                     &\phantom{=}+ \alpha \Tr(A, B)N(B) - N(B)^2,
\end{split}
\end{equation}
and hence
\begin{equation}
\begin{split}
\Delta (x)&=-\frac{4}{\alpha^2}\big\{\phantom{+\,}[\alpha\kappa+N(B)]^2\\
          &\phantom{=-\frac{4}{\alpha^2}\big(}+\big[\phantom{+\ }\alpha^3 N(A)-\alpha^2\Tr(A^\sharp, B^\sharp)\\
          &\phantom{=-\frac{4}{\alpha^2}\big(+\big[}+\alpha\Tr(A, B)N(B)-N(B)^2\big]\big\}\\
    	  &=\phantom{-}\frac{4}{\alpha^2}\{N(B^\sharp-\alpha A)-[\alpha\kappa+N(B)]^2\},
\end{split}
\end{equation}
as required. Had we started with $N(\cQ)$ we would have obtained the analogous black hole equation.

\subsection{Greatest common divisors, discrete U-duality invariants and dyon orbits}
\label{sec:gcds}

Macroscopic physical quantities, such as the leading order Bekenstein-Hawking entropy, are necessarily invariant under the continuous U-duality group of the underlying low energy supergravity action. For example, the lowest order black hole entropy of $\mathcal{N}=8$, $D=4$ supergravity is determined by the unique quartic $E_{7(7)}(\mathds{R})$ invariant $\Delta(x)$. However, in the full quantum theory this continuous symmetry is broken to a discrete subgroup due to the Dirac-Schwinger quantisation conditions. Consequently, the physical quantities of the quantised theory may also depend on a number of previously absent discrete invariants. Moreover, the U-duality charge orbits are furnished with an increased level of subtlety and their full characterisation may depend crucially on the new discrete invariants. For example, see \cite{Banerjee:2007sr} for a complete treatment of the T-duality dyon orbits of the heterotic string on a $T^6$, which depend not only on the continuous $SL(2, \mathds{R})\times SO(6,22, \mathds{R})$ quartic invariant but also on two further discrete invariants of the fully quantised U-duality group $SL(2, \mathds{Z})\times SO(6,22, \mathds{Z})$.

Typically, these discrete invariants are given by greatest common divisors of particular dyon charge combinations. As such, they are obviously not defined in the continuous case and may only be introduced for quantised charges. Accordingly, before presenting some of the key features of discrete invariants and charge orbits in $D=5$ and $D=4$,  we begin by recalling some useful properties of the greatest common divisor ($\gcd$) of integers $a,b,c$:
\begin{enumerate}
\item The $\gcd$ is commutative and associative,
    \begin{equation}
    \begin{split}
    \gcd(a,b)        &=\gcd(b,a),\\
    \gcd(\gcd(a,b),c)&=\gcd(a,\gcd(b,c))\\
                     &=\gcd(a,b,c).
    \end{split}
    \end{equation}
    \item The $\gcd$ satisfies the following basic identities,
	\begin{equation}\label{eq:gcdrelns}
	\begin{split}
	\gcd(ac,bc)&=c\gcd(a,b)\\
	\gcd(a+cb,b)&=\gcd(a,b)\\
	\gcd(b/c,a/c)&=\gcd(a,b)/c\text{\ \ for\ }c|a,b.
	\end{split}
	\end{equation}
 \end{enumerate}

\subsection*{\texorpdfstring{$D=5$:}{D=5}}
\label{sec:D5}

\begin{enumerate}
\item For an element $X$ of an integral Jordan algebra, an integer $d$ \emph{divides} $X$, denoted $d|X$, if $X=dX'$ with $X'$ integral.

\item The $\gcd$ of a collection of not all zero integral Jordan algebra elements is defined to be the greatest integer that divides them. By definition $\gcd$ is positive. The $\gcd$ may be used to define the following set of discrete U-duality invariants \cite{Krutelevich:2002}:
    \begin{equation}\label{eq:DiscreteInvariants5}
    \begin{split}
    d_1(X)&=\gcd(X)\\
    d_2(X)&=\gcd(X^\sharp)\\
    d_3(X)&=|N(X)|.
    \end{split}
    \end{equation}
\item An $n\times n$ matrix $X$ is said to be in \emph{Smith normal form} if $X$ is a diagonal matrix
    \begin{equation}\label{eq:smithnormal}
    \begin{split}
    X&=(X_1, X_2 \ldots X_n)\\
     &\equiv\diag(X_1, X_2 \ldots X_n), \  X_i \in \mathds{Z},
    \end{split}
    \end{equation}
    with $X_i|X_{i+1}$ for all $i=1,2\ldots n-1$ and all zeros lie in the bottom right corner.
\item When $\mathfrak{J}$ is $J_{3}^{\mathds{A}}$, where $\mathds{A}$ is one of the three integral split  composition algebras $\mathds{C}^s, \mathds{H}^s$ or $\mathds{O}^s$, which includes the all important $\mathcal{N}=8$ example with $E_{6(6)}(\mathds{Z})$ U-duality, the most general black string charges $A$ (or equally black hole charges $B$) may be brought into Smith normal form by a U-duality transformation
    \begin{equation}\label{eq:Smith}
    A=(A_1, A_2, A_3),
    \end{equation}
    with $A_1|A_2$, $A_2|A_3$ and $A_1, A_2\geq 0$ \cite{Krutelevich:2002}. Note, in the $\mathfrak{J}=\mathds{Z\oplus Z\oplus Z}$ case, while the charges are already in diagonal form, the reduced structure group in not large enough to put them in Smith normal form.

\item For $\mathfrak{J}=J_{3}^{\mathds{A}}$ where $\mathds{A}$ is one of the three integral split  composition algebras $\mathds{C}^s, \mathds{H}^s$ or $\mathds{O}^s$, which again includes the central $\mathcal{N}=8$ example, the orbit representatives of all black strings (holes) have been fully classified  \cite{Krutelevich:2002}. By virtue of the fact that any element $A$ is U-duality equivalent to a Smith normal form \eqref{eq:Smith} the complete set of U-duality orbit representatives may be written as:
    \begin{equation}
    k(1, l, lm), \text{\ where\ } k, l\geq 0, m\in\mathds{Z}.
    \end{equation}
    That this gives the complete set of distinct orbits follows from the fact that $k,l$ and $m$ are uniquely determined by the U-duality invariants $d_1(A)$, $d_2(A)$ and $N(A)$. This follows simply from $A=k(1, l, lm)$ with  $A^\sharp=k^2(l^2m, lm, l)$:
    \begin{equation}
    \begin{split}
    d_1(A)&=k\\
    d_2(A)&=k^2\gcd(l^2m, lm, l)=k^2l\\
    d_3(A)&=k^3l^2|m|,
    \end{split}
    \end{equation}
    so that $d_1(A)$ fixes $k$, $d_2(A)=d_1^2(A)l$ determines $l$ and then $m$ is set by $d_3(X)$ with sign given by $\sgn(N(X))$.  Consequently, the Smith normal form of any black string (hole) is unique; any two black strings $A$ and $A'$ with $d_1(A)=d_1(A')$, $d_2(A)=d_2(A')$ and $N(A)=N(A')$ are U-duality related. Conversely, two black strings with distinct Smith normal forms are not U-duality related. A simple example is given by,
    \begin{align}
    A&=k(1, 1, 1), & A'&= (1, k, k^2),
    \end{align}
    for which, $N(A)=N(A')=k^3$, but $d_2(A)=k^2$ and $d_2(A')=k$.
    \begin{quote}
    \emph{There are black string (hole) configurations with the same cubic norm and hence lowest order entropy that are not U-duality related.}
    \end{quote}

\item A black string $A$ (black hole $B$) is said to be \emph{primitive} if $d_1(A)=1$ ($d_1(B)=1$). A primitive black hole in Smith normal form clearly has $k=1$. For primitive black holes in $\mathcal{N}=8$, $D=5$ type II string theory a degeneracy counting formula has been derived in \cite{Maldacena:1999bp} which depends not only on the leading order entropy $N(B)$ but also on the discrete invariant $d_2(B)$.

\end{enumerate}

\subsection*{\texorpdfstring{$D=4$:}{D=4}}
\label{sec:D4}

\begin{enumerate}
\item For an element $x$ of an integral FTS, an integer $d$ \emph{divides} $x$, denoted $d|x$, if $x=dx'$ with $x'$ integral.
\item The $\gcd$ of a collection of not all zero integral FTS elements is defined to be the greatest integer that divides them. By definition $\gcd$ is positive. The $\gcd$ may be used to define the following\footnote{In the $\mathcal{N}=8$ case Sen \cite{Sen:2008sp} denotes  $d'_2(x)$ by $\psi $, and $d_5(x)$ by $\chi$.} set of discrete U-duality invariants \cite{Krutelevich:2004,Sen:2008sp}:
    \begin{equation}\label{eq:DiscreteInvariants}
    \begin{split}
    d_1(x)&=\gcd(x)\\
    d_2(x)&=\gcd(3\,T(x,x,y)+\{x,y\}\,x) \ \forall\ y\\
    d'_2(x)&=\gcd(\cP(x),\,\cQ(x),\, \mathcal{R}(x))\\
    d_3(x)&=\gcd( T(x,x,x) )\\
    d_4(x)&=|\Delta(x)|\\
    d_5(x)&=\gcd(x \wedge T(x)),
    \end{split}
    \end{equation}
    where $\wedge$ denotes the antisymmetric tensor product. $\cP=B^\sharp -\alpha A$ and $\cQ=A^\sharp -\beta B$ are the charge combinations appearing in the 4D/5D lift \eqref{eq:5DPQ} and   $\mathcal{R}(x):\mathfrak{J}\to \mathfrak{J}$ is a Jordan algebra endomorphism given by
    \begin{equation}
    \mathcal{R}(x)(C)=2\kappa(x)C+2\{A,B,C\},\ C\in\mathfrak{J},
    \end{equation}
    where $\{A,B,C\}$ is the Jordan triple product \eqref{eq:Jtripleproduct}. Taken together, $(\cP(x),\,\cQ(x),\, \mathcal{R}(x))$ form the adjoint representation of  the 4D U-duality: $\rep{133}$ in the case of $E_{7(7)}(\mathds{Z})$. Under the 5D U-duality, they transform as the fundamental, contragredient fundamental and adjoint representations, respectively: $\rep{27}$, $\rep{27'}$ and $\rep{1+78}$ in the case of  $E_{6(6)}(\mathds{Z})$.

    The second discrete invariant $d_2(x)$ may be rephrased using the fact that an integer $n$ divides $3\,T(x,x,y)+\{x,y\}\,x$ for all $y$ if and only if it divides the following five expressions \cite{Krutelevich:2004}:
    \begin{equation}
    2\,\cP,\ 2\,\cQ,\ 3\alpha\beta-\Tr(A, B),\ \mathcal{R}(x),\ \mathcal{R}(x'),
    \end{equation}
    where
    \begin{align}
    x&=\begin{pmatrix}\alpha&A\\B&\beta\end{pmatrix} ,& x'&=\begin{pmatrix}\beta&B\\A&\alpha\end{pmatrix}.
    \end{align}
    Note, on restricting to the $STU$ subsector with $\mathfrak{J}=\mathds{Z}\oplus\mathds{Z}\oplus\mathds{Z}$
    \begin{equation}\label{eq:RinSTU}
    \mathcal{R}(x)=(\alpha\beta-\Tr(A,B))\mathds{1}+2 A\circ B,
    \end{equation}
    and therefore $\mathcal{R}(x) = \mathcal{R}(x')$ and $3\alpha\beta-\Tr(A, B)=\Tr(\mathcal{R}(x))$ so that, using the  $STU$ notation presented in \autoref{sec:STUmodel} and \autoref{tab:Dictionaries}, one obtains
    \begin{equation}
    d_2(x)=\gcd(\gamma^{A}, \gamma^{B}, \gamma^{C}),
    \end{equation}
    with $\gamma^{A}, \gamma^{B}, \gamma^{C}$  given in \eqref{eq:ABCgammas} transforming respectively as $\rep{(3,1,1), (1,3,1), (1,1,3)}$  under $SL(2,\mathds{Z}) \times SL(2,\mathds{Z}) \times SL(2,\mathds{Z})$.
\item $x$ is said to be \emph{reduced} if it is of the form
    \begin{equation}\label{eq:reducedform}
    x=\begin{pmatrix}\alpha&A\\0&\beta\end{pmatrix},
    \end{equation}
    with $\alpha >0$, $\alpha|\beta$, $\alpha|A$. $x$ is said to be \emph{diagonal reduced} if in addition $A$ is diagonal. For a reduced $x$, $d_1(x)=\alpha$.
\item When $\mathfrak{J}$ is either $\mathds{Z}\oplus\mathds{Z}\oplus\mathds{Z}$ or $J_{3}^{\mathds{A}}$, where $\mathds{A}$ is one of the three integral split  composition algebras $\mathds{C}^s, \mathds{H}^s$ or $\mathds{O}^s$, which includes the all important $\mathcal{N}=8$ example with $E_{7(7)}(\mathds{Z})$ U-duality\footnote{But excludes the magic $\mathcal{N}=2$ supergravities based on $\mathds{C}, \mathds{H}$ or $\mathds{O}$ \cite{Gunaydin:1983bi,Gunaydin:1983rk,Gunaydin:1984ak}, which require a separate treatment \cite{Gross:1996,Krutelevich:2002}.}, the most general black hole charge $x$ may be brought by a U-duality transformation to the diagonal reduced canonical form depending on just five parameters \cite{Krutelevich:2004}
    \begin{equation}\label{eq:diagreduced}
    x=\begin{pmatrix}\alpha&(A_1,A_2,A_3)\\0&\beta\end{pmatrix},
    \end{equation}
    with $\alpha >0$, $\alpha|\beta$, $\alpha|A$.

\item Moreover, for the cases with $\mathfrak{J}=J_{3}^{\mathds{A}}$, $A$ may be transformed into \emph{Smith} diagonal form so that $A_1|A_2$, $A_2|A_3$ and $A_1, A_2\geq 0$ \cite{Krutelevich:2002,Krutelevich:2004}, in which case we may write the most general black hole in a further simplified form:
    \begin{equation}\label{eq:canonicaldiagreduced}
    x=\alpha\begin{pmatrix}1&k(1,l,lm)\\0&j\end{pmatrix},
    \end{equation}
    with $k, l\geq 0$. In this notation the discrete U-duality invariants \eqref{eq:DiscreteInvariants} are given by:
    \begin{equation}\label{eq:DiscreteInvariants_jklm}
    \begin{split}
    d_1(x)&=\alpha\\
    d_2(x)&=\alpha^2\gcd(j,2k)\\
    d'_2(x)&=\alpha^2\gcd(j,k)\\
    d_3(x)&=\alpha^3\gcd(j,2k^2l)\\
    d_4(x)&=\alpha^4|j^2+4k^3l^2m|\\
    d_5(x)&=\alpha^4\gcd(j,k^2l).
    \end{split}
    \end{equation}
    However, unlike the $D=5$ case the invariants \eqref{eq:DiscreteInvariants_jklm} are insufficient to determine uniquely $j,k,l,m$, as can be seen by taking any example with $j=1$. Note, however, that $\alpha$ is clearly fixed by $d_1(x)$.  Consequently, the reduced canonical form \eqref{eq:canonicaldiagreduced} of any given black hole is not necessarily unique and, to the best of our knowledge, there is no complete classification of the U-duality orbits. For example,
    \begin{equation}
    \begin{split}
    x &=\alpha\begin{pmatrix}1 & (0,0,0) \\ (0,0,0) & j\end{pmatrix}, \\
    x'&=\alpha\begin{pmatrix}1 & (j,0,0) \\ (0,0,0) & j\end{pmatrix},
    \end{split}
    \end{equation}
    are both in canonical form and U-duality related using $\phi(C)$ in \eqref{eq:FreudenthalConstructionTransformations} with $C=(1,0,0)$.

    It is clear from \eqref{eq:DiscreteInvariants_jklm} that there are black holes with the same quartic norm but differing discrete invariants. A simple example is given by:
    \begin{equation}
    \begin{split}
    x &=\begin{pmatrix}1&2(1,0,0)\\0&2\end{pmatrix},\\
    x'&=\begin{pmatrix}1& (1,0,0)\\0&2\end{pmatrix},
    \end{split}
    \end{equation}
    for which, $\Delta(x)=\Delta(x')=4$, but $d'_2(x)=2$ and $d'_2(x')=1$. In summary
    \begin{quote}
    \emph{There are black hole configurations with the same quartic norm and hence lowest order entropy that are definitely not U-duality related;}
    \end{quote}
    but more surprisingly:
    \begin{quote}
    \emph{There are black hole configurations having the same quartic norm and same discrete invariants of \eqref{eq:DiscreteInvariants_jklm} that are apparently not U-duality related.}
    \end{quote}
    A simple example is given by:
    \begin{equation}
    \begin{split}
    x &=\begin{pmatrix}1&(1,2,2)\\0&1\end{pmatrix},\\
    x'&=\begin{pmatrix}1&(1,1,4)\\0&1\end{pmatrix}.
    \end{split}
    \end{equation}
However, without a complete classification of the U-duality orbits one is not able to be certain in general about the U-equivalence of black hole charge vectors.

\item $x$ is said to be \emph{primitive} if $d_1(x)=1$. A primitive diagonally reduced black hole clearly has $\alpha=1$.

\item   An element $x$ is said to be  \emph{projective} if its U-duality orbit contains a diagonal reduced element satisfying
    \begin{equation} \label{eq:projective element def}
    \begin{split}
    \gcd(\alpha A_1, \alpha\beta, A_2A_3)&=1;  \\
    \gcd(\alpha A_2, \alpha\beta, A_1A_3)&=1;  \\
    \gcd(\alpha A_3, \alpha\beta, A_1A_2)&=1.  \\
    \end{split}
    \end{equation}
    The concept of a projective element was originally introduced for the case $\mathfrak{J}=\mathds{Z}\oplus\mathds{Z}\oplus\mathds{Z}$ along with certain generalisations central to the new view on Gauss composition and its extension as expounded in \cite{Bhargava:2004}. This is the definition relevant to the $STU$ model and is related to $d'_2(x)$. It is given by
    \begin{equation}
    \gcd(\cP_i(x),\,\cQ_i(x),\, \mathcal{R}_i(x))=1,\ i=1,2,3,
    \end{equation}
    where the index $i$ refers to the three components of $\mathds{Z}\oplus\mathds{Z}\oplus\mathds{Z}$ and $\mathcal{R}(x)$ is in the reduced form \eqref{eq:RinSTU}. In the $STU$ language of \autoref{sec:STUmodel} this projectivity condition is
    \begin{equation}
    \gcd(\half\gamma^{i}_{00},\, \half \gamma^{i}_{11},\, \gamma^{i}_{01})=1,\ i=A,B,C,
    \end{equation}
    where $\gamma^i$ is given in \eqref{eq:ABCgammas}. Using \eqref{eq:pqtoPQdic} this becomes
    \begin{equation}
    \gcd(\half P^{2}_i,\, \half Q^{2}_i,\, P_i\cdot Q_i)=1,\ i=A,B,C,
    \end{equation}
    where the index $i$ refers to the three triality related versions of $P, Q$ transforming as a $\rep{(2, 4)}$ of $SL_{A}(2,\mathds{Z})\times SO(2,2,\mathds{Z})$, $SL_{B}(2,\mathds{Z})\times SO(2,2,\mathds{Z})$ or $SL_{C}(2,\mathds{Z})\times SO(2,2,\mathds{Z})$ \cite{Duff:1995sm, Borsten:2008wd}. The idea was then further generalised in \cite{Krutelevich:2004} giving the appropriate definition for $\mathcal{N}=8$, $D=4$ black holes which we adopted here \eqref{eq:projective element def}.

    The class of projective FTS elements is of particular relevance to recent developments in number theory \cite{Bhargava:2004, Krutelevich:2004}.
\item If $x$ is projective then $d'_2(x)=1$.
\item If $d_3(x)=1$ then $x$ is projective.
\item If $d_3(x) \geq 3$ or $T(x)=0$ then $x$ is not projective.
\item When $\Delta$ is odd, $d_3(x)=1$ iff $x$ is projective.
\item While the general treatment of orbits in $D=4$ is lacking, the orbit representatives of  \emph{projective} black holes have been fully classified in \cite{Krutelevich:2004}, at least for $\mathfrak{J}=J_{3}^{\mathds{A}}$ where $\mathds{A}$ is one of the three integral split  composition algebras $\mathds{C}^s, \mathds{H}^s$ or $\mathds{O}^s$, which again includes the central $\mathcal{N}=8$ example.  Any projective element $x$ is U-duality equivalent to an element \cite{Krutelevich:2004}:
    \begin{equation}\label{eq:projective_fts}
    \begin{gathered}
    \begin{pmatrix}1 & (1,1,m) \\(0,0,0) &j \\ \end{pmatrix},\\j\in \{0, 1\},\ m\in\mathds{Z},
    \end{gathered}
    \end{equation}
    where the values of $m$ and $j$ are uniquely determined by $\Delta(x)$. Further,
    \begin{itemize}
\item U-duality, for example $E_{7(7)}(\mathds{Z})$, acts transitively on projective elements of a given norm $\Delta(x)$.
\item If $\Delta(x)$ is a squarefree\footnote{An integer is squarefree if its prime decomposition contains no repetition.} integer equal to 1 (mod 4) \emph{or} if $\Delta(x)=4k$, where $k$ is squarefree and equal to  $2$ or $3$ (mod 4), then $x$ is projective and hence U-duality acts transitively.
    \begin{quote}
    \emph{In the projective case all black holes with the same quartic norm and hence lowest order entropy are U-duality related.}
    \end{quote}

\end{itemize}

\end{enumerate}

\section{The 4D Freudenthal dual}
\label{sec:Fdual}
\subsection{Definition}

Given a black hole with charges $x$, we define its Freudenthal dual by
\begin{equation}\label{eq:dual}
\tilde{x}=T(x) |\Delta(x)|^{-1/2}.
\end{equation}
As described in \autoref{sec:J}, the FTS divides black holes into five distinct ranks or orbits. F-duality \eqref{eq:dual} is initially defined for large rank 4 black holes for which both $T$ and $\Delta$ are nonzero.  Small black holes are discussed in \autoref{sec:small}.

The invariance of $\Delta (x)$ follows by noting that
\begin{equation}
2\Delta(x)=\{T(x), x\}
\end{equation}
where $T(x)=T(x,x,x)$ obeys
\begin{equation}\label{eq:TofTofx}
T(T(x))=-\Delta^2(x)x
\end{equation}
and hence
\begin{equation}
\Delta(T(x))=\Delta(x)^3
\end{equation}
So
\begin{equation}
\Delta(\tilde{x})=\Delta(T(x))\Delta(x)^{-2}=\Delta(x).
\end{equation}
Moreover
\begin{equation}
\tilde{\tilde{x}}= T(\tilde{x})|\Delta(x)|^{-1/2}= T(T(x))\Delta(x)^{-2}= -x.
\end{equation}

In the case of two black holes related by Freudenthal duality, the Dirac-Schwinger quantisation condition \eqref{eq:DS} becomes
\begin{equation}
\{\tilde{x}, x \}= \{T(x),x \}|\Delta(x)|^{-1/2}=2 \sgn(\Delta)|\Delta(x)|^{1/2}
\end{equation}
which is also invariant.  Note the factor of 2.

As noted in \autoref{sec:Introduction}, for a valid dual charge vector $\tilde{x}$, we require that $|\Delta(x)|$ is a perfect square. So we may write
\begin{equation}
 |\Delta(x)|=\tfrac{1}{4}\{\tilde {x}, x \}^2
\end{equation}
with
\begin{equation}
\{\tilde{x}, x\}=\tilde {\alpha}\beta-\tilde{\beta}\alpha+\Tr(\tilde A,B)-\Tr(\tilde B,A),
\end{equation}
This is a necessary, but not sufficient condition because we further require that
\begin{equation}\label{eq:d_icondition}
d_4(x)=\left[\frac{d_3(x)}{d_1(\tilde x)} \right]^2=\left[ \frac{d_3(\tilde x)}{d_1(x)} \right]^2=d_4(\tilde x).
\end{equation}
Since F-duality requires that $\Delta(x)$ is a perfect square, the squarefree condition discussed in item 12 of \autoref{sec:gcds} does not apply to the subset of black holes admitting an F-dual, which may or may not be projective:
\begin{quote}
\emph{Non-projective black holes related by an F-duality not conserving $d_1$ provide examples of configurations with the same quartic norm and hence lowest order entropy that are definitely not U-duality related,}
\end{quote}
but more surprisingly,
\begin{quote}
\emph{Non-projective black holes related by an F-duality conserving $d_1$ provide examples of configurations with the same quartic norm, and same discrete invariants \eqref{eq:DiscreteInvariants_jklm}, that are apparently not U-duality related. Furthermore, without a complete orbit classification it is still an open question  whether such black holes in general are U-duality related \cite{Krutelevich:2004}.}
\end{quote}

The U-duality integral invariants $\{x,y\}$ and $\Delta(x,y,z,w)$ are not generally invariant under Freudenthal duality while $ \{\tilde{x},x\}$, $\Delta(x)$,  and hence the lowest-order black hole entropy, are invariant.  However, higher order corrections to the black hole entropy depend on some of the discrete U-duality invariants, to which we now turn.

\subsection{The action of F-duality on discrete U-duality invariants}
\label{sec:Fdualdiscrete}

The first important observation we make is that since
\begin{equation}
\begin{gathered}
T(\sigma(x), \sigma(y), \sigma(z))= \sigma(T(x, y, z)),\\
\forall\ \sigma\in \Aut{\mathfrak{M(J)}},
\end{gathered}
\end{equation}
F-duality commutes with U-duality
\begin{equation}
\widetilde{\sigma(x)}= \sigma(\tilde{x}).
\end{equation}
We shall see that of the discrete U-duality invariants listed in \eqref{eq:DiscreteInvariants},  not only $d_4(x)$ but also $d_2(x)$, $d'_2(x)$ and $d_5(x)$ are F-dual invariant. However,  $d_1=\gcd(x)$ and $d_3=\gcd(T(x))$ need not be.

The invariance of $d_5(x)$ follows from \eqref{eq:TofTofx} which implies
\begin{equation}
\begin{split}
\tilde{x}\wedge T(\tilde{x})&= T(x)|\Delta|^{-1/2}\wedge T(T(x)|\Delta|^{-1/2})\\
														&=-|\Delta|^{-2}T(x)\wedge \Delta^{2}x\\
														&=\sgn(\Delta)x\wedge T(x)
\end{split}
\end{equation}
and, hence, $d_5(x)=d_5(\tilde{x})$.

To prove the invariance of $d'_2(x)$, we examine $\cP(x)$, $\cQ(x)$ and $\mathcal{R}(x)$ in turn. First, for the black string magnetic charge $\cP$ we find from \eqref{eq:Tofx}
\begin{equation}\label{eq:Ptildedelta}
\begin{split}
|\Delta|\tilde{\cP}= 4\big\{&[-\alpha A^\sharp+A\times B^\sharp+\kappa(x)B]^\sharp\\
                            &-(\alpha\kappa(x)+N(B))\times\\
                            &[\phantom{-}\beta B^\sharp-B\times A^\sharp-\kappa(x)A]\big\}.
\end{split}
\end{equation}
Using $(X+Y)^\sharp=X\times Y+X^\sharp + Y^\sharp$ the first term of \eqref{eq:Ptildedelta} gives
\begin{equation}
\begin{split}\label{eq:Btildesharp}
   &[-\alpha A^\sharp+A\times B^\sharp+\kappa(x)B]^\sharp\\
=\ &-\alpha (A\times B^\sharp)\times A^\sharp+\kappa(x) (A\times B^\sharp)\times B\\
   &-\alpha\kappa(x) A^\sharp\times B+ (A\times B^\sharp)^\sharp\\
   &+\alpha^2N(A)A+\kappa(x)^2B^\sharp.
\end{split}
\end{equation}
This may be further simplified using the identities
\begin{equation}
\begin{split}
X^\sharp\times(X\times Y)\phantom{^\sharp}&=\phantom{-}N(X)Y + \Tr(X^\sharp, Y)X\\
(X\times Y)^\sharp                        &=\phantom{-}\Tr(X^\sharp, Y)Y + \Tr(Y^\sharp, X)X\\
                                          &\phantom{=\,}- X^\sharp\times Y^\sharp,
\end{split}
\end{equation}
which follow from the quadratic adjoint definition and the requirement that $(X^\sharp)^\sharp=N(X)X$.
These identities yield
\begin{equation}
\begin{split}
(A\times B^\sharp)\times A^\sharp          &=\phantom{-}N(A)B^\sharp + \Tr(A^\sharp, B^\sharp)A\\
(A\times B^\sharp)\times B\phantom{^\sharp}&=\phantom{-}N(B)A\phantom{^\sharp} + \Tr(A\phantom{^\sharp}, B\phantom{^\sharp})B^\sharp\\
(A\times B^\sharp)^\sharp\phantom{\times A\ \,}&=\phantom{-}\Tr(A^\sharp, B^\sharp)B^\sharp\\
                                           &\phantom{=\,} + \Tr(A, B)N(B)A\\
                                           &\phantom{=\,}- N(B)A^\sharp\times B.
\end{split}
\end{equation}
Using the above to simplify \eqref{eq:Btildesharp} and then substituting into  \eqref{eq:Ptilde} gives, after collecting terms,
\begin{equation}
\begin{split}
|\Delta|\tilde{\cP}&=\phantom{+}4[\alpha^2\beta - \alpha \Tr(A,B) - \alpha \kappa(x)]A^\sharp\times B\\
                   &\phantom{=}+\Delta (B^\sharp - \alpha A).
\end{split}
\end{equation}
The first term vanishes identically so that
\begin{equation}\label{eq:Ptilde}
\tilde{\cP}= \sgn(\Delta) (B^\sharp - \alpha A)=\sgn(\Delta)\cP.
\end{equation}
 A similar treatment goes through for $\cQ$:
 \begin{equation}\label{eq:Qtilde}
\tilde{\cQ}= \sgn(\Delta) (A^\sharp - \beta A)=\sgn(\Delta)\cQ.
\end{equation}
 Finally, in order to demonstrate the invariance of $\mathcal{R}(x)$ we exploit the fact that since
U-duality commutes with F-duality we may  assume  $x$ to be in reduced form \eqref{eq:reducedform} so that
\begin{equation}
\mathcal{R}(x)(C)=\alpha\beta\, C.
\end{equation}
For reduced $x$ the dual is given by
\begin{equation}
|\Delta|^{1/2}\tilde{x}=\begin{pmatrix}-\alpha^2\beta &\alpha\beta A\\2\alpha A^\sharp&\alpha\beta^2+2N(A)\end{pmatrix},
\end{equation}
where $\Delta=-\alpha^2\beta^2-4\alpha N(A)$. On substituting in for $\mathcal{R}(x)(C)$ one finds
\begin{equation}
\begin{split}
|\Delta| \mathcal{R}(\tilde{x})(C)&=\alpha\beta\big(-[\alpha^2\beta^2+8N(A)]C\\
                                  &\phantom{=\alpha\beta\big(\ }+4\alpha\{A,A^\sharp,C\}\big)\\
								  &=\Delta \alpha\beta\, C,
\end{split}
\end{equation}
where we have used $\{X,X^\sharp,Y\}=N(X)Y$ \cite{McCrimmon:2004} in the final step. Hence,  $\mathcal{R}$ is also invariant up to a sign.
\begin{equation}\label{eq:Rtilde}
\tilde{\mathcal{R}}= \sgn(\Delta) \alpha\beta C=\sgn(\Delta){\mathcal R}.
\end{equation}
This clearly establishes the invariance of $d'_2(x)$ under F-duality.

To prove the invariance of $d_2(x)$ we first rephrase the problem using the fact that an integer $n$ divides $3\,T(x,x,y)+\{x,y\}\,x$ for all $y$ if and only if it divides the following five expressions \cite{Krutelevich:2004}:
\begin{gather}
2\, \cP, \ 2\,\cQ,\ 3\alpha\beta-\Tr(A, B), \ \mathcal{R}(x),\ \mathcal{R}(x'),
\shortintertext{where}
\begin{aligned}
x&=\begin{pmatrix}\alpha&A\\B&\beta\end{pmatrix} ,& x'&=\begin{pmatrix}\beta&B\\A&\alpha\end{pmatrix}.
\end{aligned}
\end{gather}
Hence, we are only further required to establish the invariance of $3\alpha\beta-\Tr(A, B)$. The proof goes along much the same lines to obtain
\begin{equation}
3\tilde{\alpha}\tilde{\beta}-\Tr(\tilde{A}, \tilde{B})=\sgn(\Delta)[3\alpha\beta-\Tr(A, B)].
\end{equation}

Finally, recall that restricting to the $STU$ subsector $d_2(x)$ takes the reduced form
\begin{equation}
 d_2(x)=\gcd(\gamma^{A}, \gamma^{B}, \gamma^{C}).
\end{equation}
In this case the proof of F-dual invariance  is simplified since each $\gamma$ is individually invariant, up to a sign, under F-duality \eqref{eq:gammaofduala}.

As for $d_1(x)$ and $d_3(x)$, it follows from \eqref{eq:d_icondition} that their product is invariant
\begin{equation}
 d_1(x)d_3(x)=d_1(\tilde x)d_3(\tilde x)
\end{equation}
but separately they need not be. Another way to state this is that the F-dual of a primitive black hole may not itself be primitive. To see this, recall that by definition
\begin{equation}
x=d_1(x)x_0,
\end{equation}
where $x_0$ is primitive with $d_1(x_0)=1$. Hence
\begin{equation}
T(x)=d_1(x)^3T(x_0)
\end{equation}
and
\begin{equation}
\Delta(x)=d_1(x)^4\Delta(x_0).
\end{equation}
So
\begin{equation}
\tilde x=d_1(x) T(x_0) |\Delta(x_0)|^{-1/2}=d_1(x) \tilde{x}_0
\end{equation}
and
\begin{equation}
d_1(\tilde x)=d_1(x) d_1(\tilde{x}_0).
\end{equation}
Hence $d_1(x)$ is invariant if $d_1(\tilde{x}_0)=d_1(x_0) \equiv 1$, which is not necessarily so.

Typically, the literature on exact 4D black hole degeneracies \cite{Strominger:1996sh,Dijkgraaf:1996it,Maldacena:1999bp, Pioline:2005vi, Dabholkar:2005dt, Shih:2005qf, Shih:2005uc, Pioline:2006ni, Sen:2007qy, Banerjee:2007sr, Banerjee:2008pu, Banerjee:2008ri, Banerjee:2008pv, Sen:2008ht, Sen:2008ta, Sen:2008sp} deals only with primitive black holes $d_1(x)=1$. We are not required to impose this condition and generically do not do so.

In \autoref{sec:largeblacks}, we provide examples which preserve $d_1(x)$ and examples that do not. If desired, however, one might restrict the subset of black holes admitting an F-dual even further by demanding that $d_1(x)$, and hence $d_3(x)$, be conserved.

\subsection{F-dual in canonical basis}
\label{sec:candual}

Recall that, subject to the caveats in item 12 of \autoref{sec:gcds}, we may write any black hole in the diagonally reduced canonical form \eqref{eq:canonicaldiagreduced},
\begin{equation}\label{eq:canlargeblack}
x=\alpha\begin{pmatrix}1&k(1,l,lm)\\0&j\end{pmatrix},
\end{equation}
where $\alpha>0,k,l\geq0$, and $\alpha,j,k,l,m\in\mathds{Z}$. The quartic norm of this element is
\begin{equation}\label{eq:candelta}
\Delta(x)=-(j^2+4k^3l^2m)\alpha^4.
\end{equation}
For $x$ to be a rank 4 we must impose
\begin{equation}\label{eq:largecanrank4cond}
j\neq0\lor klm\neq0
\end{equation}
where $\lor$ here denotes logical disjunction. Note that in order for the charge vector to be BPS we need $\sgn(j^2+4k^3l^2m)=-1$ and hence $\sgn(m)=-1$ is a necessary (and insufficient) condition. Using \eqref{eq:candelta} and the general form for $T(x)$, we find that the general F-dual is
\begin{equation}\label{eq:generalfdual}
\begin{split}
\tilde{x}=&\ \alpha|j^2+4k^3 l^2 m|^{-1/2}\times\\
          &\begin{pmatrix}-j&jk(1,l,lm)\\2k^2 l(lm,m,1)&j^2+2k^3l^2m\end{pmatrix}.
\end{split}
\end{equation}
In order that $\tilde x$ be integer, we need to impose the following three constraints:
\begin{subequations}\label{eq:threeconditions}
\begin{align}
|j^2+4k^3 l^2 m|^{1/2}&=n_0\in\mathds{N},\label{eq:can1}\\
\alpha j/n_0&=n_1\in\mathds{Z},\label{eq:can2}\\
2k^2l\alpha/n_0&=n_2\in\mathds{N}_0,\label{eq:can3}
\end{align}
\end{subequations}
where $\sgn n_1=\sgn j$. Equation \eqref{eq:can1} forces $\Delta$ to be a perfect square, \eqref{eq:can2} then ensures that the $\tilde \alpha$ component of $\tilde{x}$ lies in $\mathds{Z}$, and \eqref{eq:can3} guarantees that the $\tilde B$ component is integral. These conditions are also sufficient to make the $\tilde A$ and $\tilde \beta$ components integer valued. The dual system then becomes
\begin{equation}\label{eq:canonicallargedual}
\tilde{x}=\begin{pmatrix}-n_1&n_1k(1,l,lm)\\n_2(lm,m,1)&n_1j+n_2klm\end{pmatrix}.
\end{equation}
The utility of this form is that all valid dual charge vectors can be specified modulo a sign by their $j,k,l,m,n_1$ and $n_2$ values. Clearly if both $n_1$ and $n_2$ vanish the entire system vanishes, failing to preserve rank. However, $n_1$ and $n_2$ can vanish separately and still leave a rank 4 system. This is to be expected since F-dual preserves $\Delta$ so that \eqref{eq:canonicallargedual} must also satisfy \eqref{eq:largecanrank4cond}, telling us that one of $n_1,n_2$ must be nonzero, given the definitions \eqref{eq:threeconditions}. As a sanity check we may evaluate the quartic form for \eqref{eq:canonicallargedual} to discover that we require
\begin{equation}\label{eq:duallargecanrank4cond}
jn_1\neq0\lor klmn_1\neq0\lor klmn_2\neq0
\end{equation}
for the dual system to be a large black hole. Satisfyingly, \eqref{eq:duallargecanrank4cond} is equal to its logical conjunction with \eqref{eq:largecanrank4cond}. Furthermore, we find
\begin{widetext}
\begin{equation}
\begin{split}
d_1(\tilde{x})&=\gcd(n_1,n_2),\\
d_2(\tilde{x})&=\gcd(2k n_1^2+2m n_2^2,2n_1 \left(j n_1+2 k l m n_2\right),-k^2 l n_1^2+j n_1 n_2+k l m n_2^2)\\
							&=\alpha^2\gcd(j,2k),\\
d'_2(\tilde{x})&=\gcd(k n_1^2+m n_2^2,n_1 \left(j n_1+2 k l m n_2\right),-k^2 l n_1^2+j n_1 n_2+k l m n_2^2)\\
							&=\alpha^2\gcd(j,k),\\
d_3(\tilde{x})&=\alpha^3n_0,\\
d_4(\tilde{x})&=d_4(x),\\
d_5(\tilde{x})&=d_5(x).\\
\end{split}
\end{equation}
\end{widetext}
As expected, \eqref{eq:d_icondition} is satisfied and
\begin{equation}
\begin{split}
d_1( \tilde x)d_3( \tilde x)&=\alpha^3n_0 \gcd(n_1,n_2)\\
                            &=\alpha^4 \gcd(j,2k^2l)\\
                            &=d_1(x)d_3(x).
\end{split}
\end{equation}

\subsection{Large black hole examples}
\label{sec:largeblacks}

Turning now to examples, we note from \eqref{eq:largecanrank4cond} that there are only three cases to consider, $j=0\,\land\,klm\neq0$, $j\neq0\land klm=0$, and the most complicated case $jklm\neq0$ (where $\land$ is logical conjunction). A number of examples satisfying these conditions and the constraints \eqref{eq:threeconditions} are specified in \autoref{tab:largeexamples}.
\begin{table*}
\caption[Large black hole F-duality in canonical basis]{Conditions on parameters for several example FTSs, where $p\in\mathds{Z}$. Parameters $n_0,n_1$ and $n_2$ are fixed by \eqref{eq:threeconditions}. Note that we still require $\alpha>0$, $k,l\geq0$ and $n_0\neq0$ in all cases.}\label{tab:largeexamples}
\begin{ruledtabular}
\begin{tabular}{cc*{9}{M{c}}c}
& Case  & j      & k      & l     & m              & \sgn\Delta & n_0         & n_1          & n_2             &\\
\hline
& 1     & 0      & p^2|m| & >0    & \neq0          & -\sgn m    & 2|p^3|m^2l  & 0            & |p|\alpha       &\\
\hline
& 2.1   & \neq0  & ~~~0   & \geq0 & \in\mathds{Z}  & -          & |j|         & \alpha\sgn j & 0               &\\
& 2.2   & \neq0  & >0     & ~~~0  & \in\mathds{Z}  & -          & |j|         & \alpha\sgn j & 0               &\\
& 2.3   & \neq0  & >0     & >0    & ~~~0           & -          & |j|         & \alpha\sgn j & 2k^2l\alpha/|j| &\\
\hline
& 3.1.1 & 2p     & 1      & 1     & -(p^2\pm1)     & \pm        & 2           & |p|\alpha    & \alpha          &\\
& 3.1.2 & 2lr    & 1      & >0    & -(r^2\pm q)    & \pm        & 2l|q|       & nr           & n               &\\
& 3.1.3 & 2lr    & >0     & >0    & 4q(q\pm r)/k^3 & -          & 2l|2q\pm r| & nr           & nk^2            &\\
& 3.2   & 2p+1   & 1      & 1     & -p(p+1)        & -          & 1           & (2p+1)\alpha & 2\alpha         &
\end{tabular}
\end{ruledtabular}
\end{table*}

\subsubsection*{\texorpdfstring{Case 1: $j=0\land klm\neq0$}{Case 1: j=0 and klm nonzero}}

We initially obtain
\begin{equation}
\begin{split}
n_0&=2l|k^3m|^{1/2},\\
n_1&=0,\\
n_2&=|k/m|^{1/2}\alpha.
\end{split}
\end{equation}
To force $n_2\in\mathds{Z}$ we make $k/m$ a perfect square: $k=p^2|m|, p\in\mathds{Z}\setminus\{0\}$. This reduces $n_0$ to $2p^3m^2l$ as given in \autoref{tab:largeexamples}. Notably, this ansatz furnishes integral $n_1$ and $n_2$ and exhausts the possibilities for the $j=0$ case. BPS and non-BPS charge vectors in this case are related by a sign flip on $m$.

\subsubsection*{\texorpdfstring{Case 2: $j\neq0\land klm=0$}{Case 2: j nonzero and klm=0}}

We immediately note that this case is, at least ostensibly, considerably more complicated than the last since there is more than one way in which $klm$ can vanish: the three ways $k,l$ and $m$ can vanish individually, the three ways they vanish in pairs, and the one way they can all vanish. Nevertheless we only need to consider three cases out of these seven since, glancing at \eqref{eq:canlargeblack}, we note that when $k=0$ the values of $l$ and $m$ are irrelevant, and similarly when $k\neq0\land l=0$ the value of $m$ is irrelevant. In all three subcases we have
\begin{equation}
\begin{split}
n_0&=|j|,\\
n_1&=\sgn j\alpha,\\
n_2&=2k^2l\alpha/|j|.
\end{split}
\end{equation}
Clearly $n_2\in\mathds{Z}$ is the problematic condition.

\begin{description}
\item[Subcase 2.1: $k=0$.] We immediately have $n_2=0$.

\item[Subcase 2.2: $l=0$.] We immediately have $n_2=0$.

\item[Subcase 2.3: $m=0$.] The remaining case sets only $m=0$ and it remains for $2k^2l\alpha/j\in\mathds{Z}$ to be imposed. In general $j$ may divide $k$ and/or $l$ and/or $\alpha$ individually and one would have to resort to a prime decomposition to progress. Further discussion of this subcase may be found in \hyperref[sec:more]{appendix A}.
\end{description}

\subsubsection*{\texorpdfstring{Case 3: $jklm\neq0$}{Case 3: jklm nonzero}}

By far the most taxing case, since the perfect square requirement in general demands the solution of the Diophantine equation $j^2+4k^3 l^2 m=\pm p^2, 0\neq p\in\mathds{Z}$. This case does however include BPS elements whereas cases 2 forced $\sgn\Delta=-1$. The examples presented in \autoref{tab:largeexamples} postulate either odd or even $j$ and restrict $k,l$ so that only $m$ needs to compensate for $j$.
\begin{description}
\item[Subcase 3.1: $j=2p$.] We have
    \begin{equation}
    \begin{gathered}
    |(2p)^2+4k^3l^2m|=(2q)^2,\ q\in\mathds{Z} \\
    \Rightarrow\ k^3l^2m=\pm q^2-p^2,
    \end{gathered}
    \end{equation}
    and it remains to find $k,l,m$ satisfying this requirement. Further discussion of this general subcase may be found in \hyperref[sec:more]{appendix A}.

    \begin{description}
    \item[3.1.1] Restricting to $k=l=1$, choose a compensating $m$:
        \begin{gather}
        m=-(p^2\pm q^2)
        \shortintertext{under which}
        \begin{split}
        n_0&=2|q|\\
        n_1&=p\alpha/|q|\\
        n_2&=\alpha/|q|.
        \end{split}
        \end{gather}
        Consequently we must restrict to $q=\pm1$. This nevertheless leaves open the possibility that the charge vector is BPS since $\sgn\Delta$ corresponds to the sign choice in the $m$ postulate.

    \item[3.1.2] If we now allow $l$ to be arbitrary and set $p=lr$, $r\in\mathds{Z}$ we can choose the same compensating $m$, but are forced to make an ansatz for $\alpha$ to make $\tilde{x}$ and $\Delta$ integer.
        \begin{gather}
        \begin{split}
        m&=-(r^2\pm q^2),\\
        \alpha &= n|q|,
        \end{split}
        \shortintertext{under which}
        \begin{split}
        n_0&=2l|q|\\
        n_1&=nr\\
        n_2&=n.
        \end{split}
        \end{gather}
        where $n\in\mathds{N}$. This subcase has the additional property that $d_1$ and $d_3$ change under F-duality (see \autoref{tab:examples_table}). Further discussion of this subcase may be found in \hyperref[sec:more]{appendix A}.

    \item[3.1.3] A final example is given by imposing $k^3m=4q(q\pm r)$ so that $\Delta(x)=-4l^2(2q\pm r)^2\alpha^4$ (here we still have $p=lr$). This yields:
        \begin{equation}
        \begin{split}
        n_0&=2l|2r\pm q|\\
        n_1&=nr\\
        n_2&=nk^2,
        \end{split}
        \end{equation}
        where clearly we have made the further imposition $\alpha=n|2q\pm r|$. Hence, for $r=1$, $d_1(\tilde{x})=n$ so that we may take a non-primitive black hole to a primitive F-dual black hole.
    \end{description}

\item[Subcase 3.2: $j=2p+1$.] In contrast to the previous subcase, $m$ must now counter a linear term in $j$:
    \begin{gather}
    m=-p(p+1)
    \shortintertext{under which}
    \begin{split}
    n_0&=1\\
    n_1&=(2p+1)\alpha\\
    n_2&=2\alpha.
    \end{split}
    \end{gather}
    This sets $\Delta$ to $-\alpha^4$ so all such black holes are non BPS.

\end{description}

A summary of the examples considered, along with explicit forms of $x$ and $\tilde{x}$ are presented in \autoref{tab:examples_table}.

\subsubsection*{Projective canonical black holes}

Since the projective black holes have been fully classified \eqref{eq:projective_fts}, the complete subset admitting an F-dual may be computed in a concise manner which we may therefore present in full. Recall, all projective black holes are U-dual to the projective canonical form,
\begin{gather}
x_{\text{proj}}=\begin{pmatrix}1 & (1,1,m) \\(0,0,0) & j\end{pmatrix}
\shortintertext{with quartic norm}
\Delta(x_{\text{proj}})=-(j^2+4m)
\end{gather}
where $m\in\mathds{Z}$ and $j\in\{0,1\}$ are uniquely determined from the starting FTS $x$ by
\begin{equation}
\begin{split}
m&\equiv\tfrac{1}{4}(\Delta(x)-j^2)\\
j&\equiv\Delta(x)\mod2.
\end{split}
\end{equation}
Clearly $x_{\text{proj}}$ is obtained from \eqref{eq:canlargeblack} by setting $\alpha$, $k$, and $l$ to unity, so we may carry over the results of our previous analysis. The dual charge vector is
\begin{gather}
\tilde{x}_{\text{proj}}=\begin{pmatrix}-n_1&n_1(1,1,m)\\n_2(m,m,1)&n_1j+n_2m\end{pmatrix},
\shortintertext{the preserved discrete invariants are}
\begin{array}{r@{\ =\ }c@{\ =\ }l}
d_2(\tilde{x}_{\text{proj}})         & d_2(x_{\text{proj}})  & \gcd(2,j),\\
d'_2(\tilde{x}_{\text{proj}})        & d'_2(x_{\text{proj}}) & 1,\\
d_4(\tilde{x}_{\text{proj}})         & d_4(x_{\text{proj}})  & n_0{}^2,\\
\chi(\tilde{x}_{\text{proj}})        & \chi(x_{\text{proj}}) & 1,
\end{array}
\shortintertext{while the altered discrete invariants are}
\begin{aligned}
d_1(x_{\text{proj}})&=1&d_1(\tilde{x}_{\text{proj}})&=\gcd(n_1,n_2)\\
d_3(x_{\text{proj}})&=\gcd(2,j)&d_3(\tilde{x}_{\text{proj}})&=n_0,
\end{aligned}
\shortintertext{where the $n_i$ are now given by}
\begin{split}
n_0&=|j^2+4m|^{1/2}\\
n_1&=j/n_0\\
n_2&=2/n_0.
\end{split}
\end{gather}
There are only two possible cases:
\begin{description}
\item[Case P.1: $j=0$.] The $n_i$ simplify to
    \begin{equation}
    \begin{split}
    n_0&=2|m|^{1/2}\\
    n_1&=0\\
    n_2&= |m|^{-1/2},
    \end{split}
    \end{equation}
    which clearly requires $m=\pm 1$ for integral $\tilde{x}$. This is evidently subcase 3.1 with $p=0$. See \autoref{tab:examples_table} example P.1 for the explicit form of the $x$ and $\tilde{x}$.

    This example is just the primitive Reissner-Nordstrom rank 4 black hole which may be regarded as a bound state at threshold of four singly charged primitive rank 1 black holes \cite{Duff:1994jr,Duff:1995sm,Duff:1996qp}.
\item[Case P.2: $j=1$.] We find
    \begin{equation}
    \begin{split}
    n_0&=|4m+1|^{1/2}\\
    n_1&=|4m+1|^{-1/2}\\
    n_2&= 2|4m+1|^{-1/2},
    \end{split}
    \end{equation}
    which clearly requires $m=0$ for integral $\tilde{x}$. This is evidently subcase 3.2 with $p=0$. See \autoref{tab:examples_table} example P.2 for the explicit form of the $x$ and $\tilde{x}$.
\end{description}
Recall, all projective black holes with a given norm are U-duality related. In particular, since F-duality has preserved projectivity in the above examples, the F-dual charge vector $\tilde{x}$ is necessarily U-duality related to the original $x$ as is discussed in \hyperref[sec:FDualUDualUndo]{appendix C}.

\newcolumntype{C}[1]{>{\centering} m{#1}}
\newcolumntype{X}[1]{>{\centering $} m{#1}<{$}}

\renewcommand\multirowsetup{\centering}

\newlength\adj
\setlength\adj{2mm}

\newlength\col
\setlength\col{2.3cm}

\newlength\colw
\setlength\colw{1.5cm}

\begin{turnpage}
\begin{table*}
\caption[Explicit black hole examples and their duals]{Table containing various valid integral $x$ and $\tilde{x}$. Note from the final two columns that the $U$-duality invariants $d_1$ and $d_3$ may or may not be separately conserved under F-Duality but their product always is.}\label{tab:examples_table}
\begin{ruledtabular}
\begin{tabular}{cC{0.2cm}C{\col}|M{c}M{r}| X{\colw}  *{3}{X{\colw}}|X{\colw}X{\colw} c}
\multicolumn{3}{c|}{Case}                                                                                   &     \multicolumn{2}{c|}{FTS}                                                                                                      & \Delta                                                                & d_2                                                       & d_2'                                                      & d_5                                                           & d_1                                   & d_3                                                                 &\\
\hline
&  \multirow{2}{*}[-\adj]{1}    & \multirow{2}{\col}[-\adj]{$j=0$,\\ $k=p^2\abs{m}$ }                       & x=           & \alpha\begin{pmatrix} 1 &p^2\abs{m}(1,l,lm) \\ 0,0,0 & 0\end{pmatrix}                                              &  \multirow{2}{\colw}[-\adj]{$\sgn(-m)\cdot$\\ $4p^6\alpha^4l^2m^4$}   & \multirow{2}{*}[-\adj]{$2p^2\alpha^2\abs{m}$}             & \multirow{2}{*}[-\adj]{$p^2\alpha^2\abs{m}$}              & \multirow{2}{\colw}[-\adj]{$p^4\alpha^4lm^2$}                 &      \alpha                           & 2p^4\alpha^3 l m^2                                                  &\\
&                               &                                                                           & \tilde{x} =  &  \alpha \abs{p} \begin{pmatrix}   0    &  0,0, 0\\ lm,m,1 &  p^2\abs{m}lm\end{pmatrix}                             &                                                                       &                                                           &                                                           &                                                               & \abs{p}\alpha                         & 2\abs{p} ^3\alpha^3 l m^2                                           &\\
\hline
&  \multirow{2}{*}[-\adj]{2.1}  & \multirow{2}{*}[-\adj]{$k=0$}                                             & x=           & \alpha\begin{pmatrix}1 & 0,0,0    \\  0,0,0 & j\end{pmatrix}                                                       &  \multirow{2}{\colw}[-\adj]{$-\alpha^4j^2$}                           & \multirow{2}{\colw}[-\adj]{$\alpha^2\abs{j}$}             & \multirow{2}{\colw}[-\adj]{$\alpha^2\abs{j}$}             & \multirow{2}{\colw}[-\adj]{$\alpha^4\abs{j}$}                 & \multirow{2}{*}[-\adj]{$\alpha$}      & \multirow{2}{*}[-\adj]{$\alpha^3\abs{j}$}                           &\\
&                               &                                                                           & \tilde{x} =  &  \sgn(j) \alpha  \begin{pmatrix}   -1      & 0,0,0 \\  0,0,0 & j\end{pmatrix}                                      &                                                                       &                                                           &                                                           &                                                               &                                       &                                                                     &\\
\hline
&  \multirow{2}{*}[-\adj]{2.2}  & \multirow{2}{*}[-\adj]{$l=0$}                                             & x=           &  \alpha\begin{pmatrix}1 & k,0,0    \\  0,0,0 & j\end{pmatrix}                                                      &  \multirow{2}{\colw}[-\adj]{$-\alpha^4j^2$}                           & \multirow{2}{\colw}[-\adj]{$\alpha^2\cdot$\\$\gcd(j,2k)$} & \multirow{2}{\colw}[-\adj]{$\alpha^2\cdot$\\$\gcd(j,k)$}  & \multirow{2}{\colw}[-\adj]{$\alpha^4\abs{j}$}                 & \multirow{2}{*}[-\adj]{$\alpha$}      & \multirow{2}{*}[-\adj]{$\alpha^3\abs{j}$}                           &\\
&                               &                                                                           & \tilde{x} =  &   \sgn(j) \alpha  \begin{pmatrix}   -1      & k,0,0 \\  0,0,0 & j\end{pmatrix}                                     &                                                                       &                                                           &                                                           &                                                               &                                       &                                                                     &\\
\hline
&  \multirow{2}{*}[-\adj]{2.3}  & \multirow{2}{\col}[-\adj]{$m=0$,\\$j|2k^2l\alpha$}                        & x=           &  \alpha\begin{pmatrix}1 & k,kl,0   \\  0,0,0 & j\end{pmatrix}                                                      &  \multirow{2}{\colw}[-\adj]{$-\alpha^4j^2$}                           & \multirow{2}{\colw}[-\adj]{$\alpha^2$\\$\gcd(j,2k)$}      & \multirow{2}{\colw}[-\adj]{$\alpha^2$\\$\gcd(j,k)$}       & \multirow{2}{\colw}[-\adj]{$\alpha^4\cdot$\\$\gcd(j,k^2l)$}   & \alpha                                &  \alpha^3\cdot\gcd(j,2k^2l)                                         &\\
&                               &                                                                           & \tilde{x} =  & \frac{\alpha}{\abs{j}} \begin{pmatrix}   -j      & jk(1,l,0) \\  0,0,2k^2l  & j^2\end{pmatrix}                     &                                                                       &                                                           &                                                           &                                                               & \gcd(\alpha,\frac{2k^2l\alpha}{j})    &  \alpha^3\abs{j}                                                    &\\
\hline
&\multirow{2}{*}[-\adj]{3.1.1}  & \multirow{2}{\col}{$j=2p$,\\$k=l=1$,\\$m=-(p^2\pm1)$ }                    & x=           &  \alpha\begin{pmatrix}1 & 1,1,-p^2-1 \\ 0,0,0 & 2p\end{pmatrix}                                                    &  \multirow{2}{\colw}[-\adj]{$\pm 4\alpha^4$}                          & \multirow{2}{\colw}[-\adj]{$2\alpha^2$}                   & \multirow{2}{\colw}[-\adj]{$\alpha^2$}                    & \multirow{2}{\colw}[-\adj]{$\alpha^4$}                        & \multirow{2}{*}[-\adj]{$\alpha$}      & \multirow{2}{*}[-\adj]{$2\alpha^3$}                                 &\\
&                               &                                                                           & \tilde{x} =  &     \alpha\begin{pmatrix}   -p      & p(1,1,m) \\ m,m,1  & p^2\mp1\end{pmatrix}                                    &                                                                       &                                                           &                                                           &                                                               &                                       &                                                                     &\\
\hline
& \multirow{2}{*}[-\adj]{3.1.2} & \multirow{2}{\col}{$j=2lr,k=1$,\\$\alpha=\abs{q}n$,\\$m=-(r^2\pm q^2)$ }  & x=           &  \abs{q}n\begin{pmatrix}1 &1,l,lm \\ 0,0,0 & 2lr\end{pmatrix}                                                      &  \multirow{2}{\colw}[-\adj]{$\pm4q^6n^4l^2$}                          & \multirow{2}{\colw}[-\adj]{$2q^2n^2$}                     & \multirow{2}{\colw}[-\adj]{$q^2n^2$}                      & \multirow{2}{\colw}[-\adj]{$q^4n^4l$}                         & \abs{q}n                              & 2\abs{q}^3n^3l                                                   &\\
&                               &                                                                           & \tilde{x} =  &    n \begin{pmatrix}   -r      & r(1,l,lm) \\ lm,m,1   & l(r^2\mp q^2) \end{pmatrix}                               &                                                                       &                                                           &                                                           &                                                               &  n                                    &  2q^4n^3l                                                           &\\
\hline
& \multirow{2}{*}[-\adj]{3.1.3} & \multirow{2}{\col}{$j=2lr,k^3m=4q(q\pm r)$,\\$\alpha=n\abs{2q \pm r}$ }   & x=           &  n\abs{2q\pm r}\begin{pmatrix}1 &k(1,l,lm) \\ 0,0,0 & 2lr\end{pmatrix}                                             &  \multirow{2}{\colw}[-\adj]{$-4l^2\cdot$\\$(2q\pm r)^2\alpha^4$}      & \multirow{2}{\colw}[-\adj]{$2\alpha^2\cdot$\\$\gcd(lr,k)$}& \multirow{2}{\colw}[-\adj]{$\alpha^2\cdot$\\$\gcd(2lr,k)$}& \multirow{2}{\colw}[-\adj]{$\alpha^4l\cdot$\\$\gcd(2r,k^2)$}  & n\abs{2q\pm r}    			        & 2ln^3\gcd(r,k^2) \cdot \abs{2q\pm r}^3                              &\\
&                               &                                                                           & \tilde{x} =  &    n \begin{pmatrix}   -r      & rk(1,l,lm) \\ k^2(lm,m,1)   & 2lr^2+k^3lm \end{pmatrix}                           &                                                                	    &                                                           &                                                           &                                                               &  n\gcd(r,k^2)                         &  2ln^3\cdot(2q\pm r)^4                                              &\\
\hline
&  \multirow{2}{*}[-\adj]{3.2}  & \multirow{2}{\col}{$j=2p+1$,\\$k=l=1$,\\$m=-(p^2+p)$ }                    & x=           &  \alpha\begin{pmatrix}1 &1,1,-p^2-p \\ 0,0,0 & 2p + 1\end{pmatrix}                                                 &  \multirow{2}{\colw}[-\adj]{$ -\alpha^4     $}                        & \multirow{2}{\colw}[-\adj]{$\alpha^2$}                    & \multirow{2}{\colw}[-\adj]{$\alpha^2$}                    & \multirow{2}{\colw}[-\adj]{$\alpha^4$}                        & \multirow{2}{*}[-\adj]{$\alpha$}      & \multirow{2}{*}[-\adj]{$\alpha^3$}                                  &\\
&                               &                                                                           & \tilde{x} =  &   \alpha\begin{pmatrix}   -(2p+1)      & (2p+1)(1,1, m)\\ 2(m,m,1)  & 2p^2+2p+1\end{pmatrix}                       &                                                                       &                                                           &                                                           &                                                               &                                       &                                                                     &\\
\hline
& \multirow{2}{*}[-\adj]{P.1}   & \multirow{2}{\col}{$\alpha,k,l=1$,\\$m = \pm 1$,\\$j=0$}                  & x=           & \begin{pmatrix}1 & 1,1,\pm 1 \\ 0,0,0 & 0 \end{pmatrix}                                                            &  \multirow{2}{\colw}[-\adj]{$\mp4$}                                   & \multirow{2}{\colw}[-\adj]{$2$}                           & \multirow{2}{\colw}[-\adj]{$1$}                           & \multirow{2}{\colw}[-\adj]{$1$}                               & \multirow{2}{*}[-\adj]{$1$}           & \multirow{2}{*}[-\adj]{$2$}                                         &\\
&                               &                                                                           & \tilde{x} =  &  \begin{pmatrix}  0     &  0,0,0  \\ \pm1,\pm1,1 &  \pm 1 \end{pmatrix}                                            &                                                                       &                                                           &                                                           &                                                               &                                       &                                                                     &\\
\hline
&  \multirow{2}{*}[-\adj]{P.2}  & \multirow{2}{\col}[-\adj]{$\alpha,k,l,j = 1$,\\$m = 0$}                   & x=           & \begin{pmatrix}1 & 1,1,0 \\           0,0,0 & 1 \end{pmatrix}                                                      &  \multirow{2}{\colw}[-\adj]{$-1$}                                     & \multirow{2}{\colw}[-\adj]{$1$}                           & \multirow{2}{\colw}[-\adj]{$1$}                           & \multirow{2}{\colw}[-\adj]{$1$}                               &\multirow{2}{*}[-\adj]{$1$}            & \multirow{2}{*}[-\adj]{$1$}                                         &\\
&                               &                                                                           & \tilde{x} =  &  \begin{pmatrix}   -1      &  1,1,0\\ 0,0,2 & 1\end{pmatrix}                                                       &                                                                       &                                                           &                                                           &                                                               &                                       &                                                                     &
\end{tabular}
\end{ruledtabular}
\end{table*}
\end{turnpage}

\renewcommand\multirowsetup{\raggedright}

\subsection{Small black holes?}
\label{sec:small}

For large black holes there is an unambiguous F-dual stemming from the fact that both $T(x)$ and $\Delta(x)$ are nonzero. For rank 3 one finds $\Delta(x)=0$ but $T(x)\neq0$ and one would not expect an F-dual to exist. For lower ranks both quantities vanish and since $\tilde{x}$ is a vanishing cubic quantity over the square root of a vanishing quartic quantity, one might expect a finite result.

Consider Case 1 of \autoref{tab:examples_table} and put $l=0$;
\begin{equation}
\begin{split}
x        &=\phantom{|p|}\alpha\begin{pmatrix}1&p^2|m|(1,0,0)\\0&0\end{pmatrix},\\
\tilde{x}&=|p|\alpha\begin{pmatrix}0&(0,0,0)\\(0,m,1)&0\end{pmatrix}.
\end{split}
\end{equation}
This black hole and its F-dual have vanishing $T(x)$ and $\Delta(x)$; both are rank 2 according to the classification of \autoref{tab:FTSrank}.

Now consider Case 2.2 of \autoref{tab:examples_table} and put $j=0$, $k=p^2|m|$:
\begin{equation}
\begin{split}
x        &=\alpha\begin{pmatrix}1&p^2|m|(1,0,0)\\0&0\end{pmatrix},\\
\tilde{x}&=\alpha\begin{pmatrix}-1&p^2|m|(1,0,0)\\(0,0,0)&0\end{pmatrix}.
\end{split}
\end{equation}
The starting point is the same as Case 1, but the dual is different, although still rank 2.
So the F-dual depends on the order of the two operations (i) $j=0$, $k=p^2|m|$ (ii) $l=0$.

If we further set $p=0$ in Case 1
\begin{equation}
\begin{split}
x        &=\alpha\begin{pmatrix}1&(0,0,0)\\0&0\end{pmatrix},\\
\tilde{x}&=0.
\end{split}
\end{equation}
then not even the rank is conserved since $x$ is rank 1 and  $\tilde x$ is rank 0.

For Case 2.2, on the other hand,
\begin{equation}
\begin{split}
x        &=\alpha\begin{pmatrix}1&(0,0,0)\\0&0\end{pmatrix},\\
\tilde{x}&=\alpha\begin{pmatrix}-1&(0,0,0)\\(0,0,0)&0\end{pmatrix}.
\end{split}
\end{equation}
both  $x$ and  $\tilde x$ are rank 1.

In view of this apparent lack of uniqueness we shall continue to restrict the definition of F-duality to large black holes.

\section{The NS-NS sector}
\label{sec:NS}

\subsection{\texorpdfstring{$P, Q$ notation}{P, Q notation}}
\label{sec:PQBasis}

Under the decomposition of the $\mathcal{N}=8$ U-duality group $E_{7(7)}(\mathds{Z})$ to the S-duality group $SL(2,\mathds{Z})$ and the T-duality group $SO(6,6; \mathds{Z})$
\begin{equation}
E_{7(7)}(\mathds{Z}) \supset SL(2,\mathds{Z})\times SO(6,6; \mathds{Z})
\end{equation}
the $\rep{56}$ decomposes as
\begin{equation}
\rep{56\to(2,12)+(1,32)}.
\end{equation}
The $\mathbf{(2,12)}$ is identified as the NS-NS sector where as the $\mathbf{(1,32)}$ is associated with the R-R charges. Since any $\mathcal{N}=8$ charge vector $x$ is U-dual to a diagonal reduced form \eqref{eq:diagreduced}, the R-R charges can always be transformed away for a generic black hole\footnote{Answering in the affirmative the question posed in \cite{Sen:2008sp}: Can one always assume that a $D=4$, $\mathcal{N}=8$ black hole is U-duality related to a configuration with only NS-NS charges present?} and we are free to consider those black holes with only NS-NS charges present. We write the 12 electric and 12 magnetic charges as $Q$ and $P$ respectively. In this case the quartic norm takes the simple, manifestly $SL(2,\mathds{Z})\times SO(6,6; \mathds{Z})$ invariant form
\begin{equation}
\label{eq:PQdelta}
\Delta(P,Q)= P^2 Q^2-(P \cdot Q)^2.
\end{equation}
Applying the trilinear map to $x$ in this sector one finds\footnote{This form of the trilinear map also appears in \cite{Sen:2008sp}.}
\begin{equation}
\label{eq:T}
\begin{pmatrix}T_P\\ T_Q\end{pmatrix} = \begin{pmatrix} P\cdot Q   &  -P^2   \\
                                                        Q^2        &-P\cdot Q
                                        \end{pmatrix} \begin{pmatrix} P \\ Q\end{pmatrix},
\end{equation}
where $T_P$ and $T_Q$ denote the new $P$ and $Q$ components. The Freudenthal dual then becomes
\begin{equation}\label{eq:FdualNSNS}
\begin{split}
\begin{pmatrix}\tilde P\\ \tilde Q\end{pmatrix}&= \frac{1}{\sqrt{\abs{\Delta}}} \begin{pmatrix}T_P\\ T_Q\end{pmatrix}\\
                                               &=\frac{1}{\sqrt{\abs{\Delta}}}\begin{pmatrix}P\cdot Q&-P^2\\Q^2&-P\cdot Q\end{pmatrix}
\begin{pmatrix}P\\ Q\end{pmatrix}.
\end{split}
\end{equation}

While we have been focusing here on the NS-NS sector of the $\mathcal{N}=8$ theory, the same formulae \eqref{eq:PQdelta}, \eqref{eq:T}, \eqref{eq:FdualNSNS} also apply to the toroidal compactification of the heterotic string with $\mathcal{N}=4$ supersymmetry and  $SL(2,\mathds{Z})\times SO(6,22; \mathds{Z})$ U-duality. The relevant Jordan algebra is $\mathds{Z}\oplus Q_{5,21}$  \cite{Pioline:2006ni,Gunaydin:2009dq} and $P$ and $Q$ are now 28-vectors\footnote{This case is mentioned in the mathematical literature e.g.\ \cite{Krutelevich:2004} but it is not clear how many of the results of \autoref{sec:gcds} continue to apply. See however \cite{Banerjee:2007sr}.}.

In this case we may introduce a further discrete U-duality invariant, the \emph{torsion} \cite{Dabholkar:2007vk}:
\begin{equation}
r(P,Q)=\gcd(P_{\mu\nu}),
\end{equation}
where
\begin{equation}
P_{\mu\nu}=P_{\mu}Q_{\nu}-P_{\nu}Q_{\mu}.
\end{equation}

For primitive $P$ and $Q$, the complete set of independent T-duality invariants was determined in \cite{Banerjee:2007sr}. It consists of the three familiar invariants $P^2$, $Q^2$ and $P\cdot Q$, the torsion $r(P,Q)$ and two further interdependent discrete invariants $u_1$ and $u_2$ which are constructed below. If $P$ and $Q$ are not individually primitive there are two additional T-duality invariants given by $\gcd(P)$ and $\gcd(Q)$\footnote{For the heterotic string we have a complete set of T-duality invariants which uniquely determine the black hole charges up to T-duality. This contrasts with the $\mathcal{N}=8$ case and its U-duality invariants. }. Assume $P$ and $Q$ individually primitive and let $a, b$ be two charge vectors satisfying
\begin{equation}
a\cdot Q=1,\qquad b\cdot P=1.
\end{equation}
Define
\begin{equation}
\begin{split}
u_1&=a\cdot P\mod r(P,Q),\\
u_2&=b\cdot Q\mod r(P,Q).
\end{split}
\end{equation}
It was shown in \cite{Banerjee:2007sr} that $u_1, u_2$, so defined, are independent of the choice of $a, b$, are T-duality invariant and that $u_2$ is uniquely determined by $u_1$ (and vice versa). Any two such dyons are T-duality related if and only if all five invariants have identical values.

Let us consider the action of F-duality on these T-duality invariants.  $P^2$, $Q^2$ and $P\cdot Q$ are invariant up to a sign determined by the quartic norm,
\begin{equation}
\label{eq:hetF}
\begin{split}
\tilde{P}^2&=\sgn(\Delta)P^2\\
\tilde{Q}^2&=\sgn(\Delta)Q^2\\
\tilde{P}\cdot \tilde{Q}&=\sgn(\Delta)P\cdot Q.
\end{split}
\end{equation}
Moreover
\begin{equation}\label{eq:PdotQtilde}
P\cdot\tilde{Q} =-\tilde{P}\cdot Q=\sgn(\Delta )|\Delta|^{1/2},
\end{equation}
and the quantisation rule is
\begin{equation}
P\cdot\tilde{Q}- \tilde{P}\cdot Q=\sgn(\Delta )2|\Delta|^{1/2}.
\end{equation}
Note also that
\begin{equation}
\tilde{P}_{\mu\nu}=\sgn(\Delta)P_{\mu\nu}
\end{equation}
and therefore the torsion is also invariant under F-duality.

Clearly, when $d_1(P,Q)$ and $d_3(P,Q)$ are not conserved under F-duality, then neither $u_1$ nor $\gcd(P), \gcd(Q)$ are preserved . However, in  cases when $\gcd(P)=1$ and $\gcd(Q)=1$ are in fact conserved under F-duality  it is not difficult to verify that $u_1(P,Q)$ is also preserved.

Consequently, two $1/4$-BPS ($\Delta>0$) F-dual states are T-dual if and only if both $\gcd(P)=1$ and $\gcd(Q)=1$ are preserved. On the other hand, non-BPS ($\Delta<0$) F-dual states cannot be T-duality related. Moreover, since $d_1(P,Q)$ is not necessarily invariant under F-duality,  $\gcd(P)$ and $\gcd(Q)$ are not generically invariant.

It is worth emphasising that the F-duality \eqref{eq:FdualNSNS} is not generically an $SL(2,\mathds{Z})$ S-duality, but in certain specific circumstances with $\Delta$ positive the two may coincide.

\subsection{F-dual in Sen basis}
\label{sec:Sen}

Although the canonical basis of \autoref{sec:candual} is most convenient for our purposes, it is also useful to re-express our results in the basis used by Sen and collaborators \cite{Banerjee:2007sr, Banerjee:2008ri, Sen:2008sp}, which may be more familiar to the black hole community:
\begin{equation}
P=\begin{pmatrix}Q_1\\J\\Q_5\\0\end{pmatrix},\quad Q =\begin{pmatrix}0\\n\\0\\1\end{pmatrix},
\end{equation}
with $Q_5|J, Q_1$. Here, $n$ represents an NS 5-brane winding charge, $Q_1$ a fundamental string winding charge, while $J$ and $Q_5$ are units of KK monopole charge associated with two distinct circles of the $T^6$. In FTS language we have using \eqref{eq:PQtoSen}
\begin{equation}\label{eq:sengeneral}
\begin{split}
x&=\begin{pmatrix}\alpha & (A_1,A_2,A_3) \\ (B_1,B_2,B_3) & \beta \end{pmatrix}\\
 &=\begin{pmatrix}-1 & (n, Q_1, Q_5) \\ (0, 0, 0) & J\end{pmatrix},
\end{split}
\end{equation}
which is also summarised in \autoref{tab:Dictionaries}. We see immediately that $x$ is chosen to be primitive and that we must impose
\begin{equation}\label{eq:senrank4cond}
Q_5\neq0 \land (J\neq0\lor nQ_1\neq0),
\end{equation}
for $x$ to be a valid rank 4 charge vector. Using the metric
\begin{gather}
\begin{pmatrix}0&\mathds{1}_2\\\mathds{1}_2&0\end{pmatrix}
\shortintertext{we have}
Q^2=2n,\quad P^2=2Q_1Q_5,\quad P\cdot Q=J,\\
\begin{pmatrix}T_P\\ T_Q\end{pmatrix} =\begin{pmatrix}J&-2Q_1Q_5\\2n&-J\end{pmatrix},
\begin{pmatrix}P\\ Q\end{pmatrix},
\shortintertext{and}
\Delta=4nQ_1Q_5-J^2.
\end{gather}
The Freudenthal dual is then given by
\begin{equation}
\begin{split}
\tilde{P}&=|4nQ_1Q_5-J^2|^{-1/2}\begin{pmatrix}JQ_1\\J^2-2nQ_1Q_5\\JQ_5\\-2Q_1Q_5\end{pmatrix} ,\\
\tilde{Q}&=|4nQ_1Q_5-J^2|^{-1/2}\begin{pmatrix}2nQ_1\\nJ\\2nQ_5\\-J\end{pmatrix}.
\end{split}
\end{equation}
Since \eqref{eq:sengeneral} is not in the canonical form of \autoref{sec:candual} we must begin a new analysis to restrict the charges such that they provide an integer valued dual. Since $Q_5|J, Q_1$ we can write $J=s_1Q_5$ and $Q_1=s_2Q_5$, $s_i\in\mathds{Z}$ giving
\begin{gather}
P=Q_5\begin{pmatrix}s_2\\s_1\\1\\0\end{pmatrix},\quad Q =\begin{pmatrix}0\\n\\0\\1\end{pmatrix},\\
Q^2=2n,\quad P^2=2s_2Q_5{}^2,\quad P\cdot Q=s_1,\\
\begin{pmatrix}T_P\\ T_Q\end{pmatrix} =\begin{pmatrix}Q_5s_1&-2s_2Q_5{}^2\\2n&-s_1Q_5\end{pmatrix}\begin{pmatrix}P\\ Q\end{pmatrix},\\
\Delta=-(s_1{}^2-4ns_2)Q_5{}^2,
\shortintertext{so that}
\begin{split}\label{eq:PQsenfdual}
\tilde{P}&=|Q_5||s_1{}^2-2ns_2|^{-1/2}\begin{pmatrix}s_1s_2\\s_1{}^2-2ns_2\\s_1\\-2s_2\end{pmatrix},\\
\tilde{Q}&=\sgn(Q_5)|s_1{}^2-2ns_2|^{-1/2}\begin{pmatrix}2ns_2\\ns_1\\2n\\-s_1\end{pmatrix},
\end{split}
\end{gather}
whose corresponding $x$ and $\tilde{x}$ FTS charge vectors are
\begin{align}
x        &=\begin{pmatrix}-1&(n,Q_1Q_5s_2,Q_5)\\0&Q_5s_1\end{pmatrix},\\
\tilde{x}&=-\sgn(Q_5)|s_1^2-4ns_2|^{-1/2}\times\nonumber\\
         &\phantom{-}\begin{pmatrix}s_1&s_1(n,s_2Q_5,Q_5)\\2(s_2Q_5,n,ns_2)&(s_1{}^2-2ns_2)Q_5\end{pmatrix}.\label{eq:sengeneralfdual}
\end{align}
We see that in order to restrict \eqref{eq:PQsenfdual} or \eqref{eq:sengeneralfdual} to be a valid set of charges we must impose the following three constraints
\begin{subequations}\label{eq:senconditions}
\begin{align}
|s_1^2-4ns_2|^{1/2}&=k_0\in\mathds{N},\label{eq:sencondition1}\\
s_1/k_0&=k_1\in\mathds{Z},\label{eq:sencondition2}\\
k_3/k_0&=k_2\in\mathds{N}_0,\label{eq:sencondition3}
\end{align}
\end{subequations}
where $\sgn k_1=\sgn s_1$ and $k_3$ is defined by
\begin{equation}\label{eq:sendef}
k_3:=\begin{cases}k_0|k_1|&n,s_2=0\\2\gcd(n,s_2Q_5)&\text{else},\end{cases}
\end{equation}
with this definition being motivated by the relations \eqref{eq:gcdrelns}. \autoref{eq:sencondition1} furnishes a perfect square $\Delta$, while \eqref{eq:sencondition2} and \eqref{eq:sencondition3} respectively make $\tilde{P}^{0,2}$ and $\tilde{Q}^{1,3}$ (or the $\alpha$ and $B$ components of $\tilde{x}$) integral. As was the case in \autoref{sec:largeblacks}, these conditions suffice to make $\tilde{P}^{1,3}$ and $\tilde{Q}^{0,2}$ (or the $A$ and $\beta$ components of $\tilde{x}$) integral as well. The constraints \eqref{eq:senconditions} are directly analogous to \eqref{eq:threeconditions}, with the added complication of $k_3$ arising from the $A$ component of $x$ no longer being in Smith normal form.

The dual system becomes
\begin{gather}
\begin{split}
\tilde{P}&=|Q_5|\begin{pmatrix}k_1s_2\\k_1s_1-2ns_2k_2/k_3\\k_1\\-2s_2k_2/k_3\end{pmatrix},\\
\tilde{Q}&=\sgn(Q_5)\begin{pmatrix}2ns_2k_2/k_3\\nk_1\\2nk_2/k_3\\-k_1\end{pmatrix},
\end{split}\\
\begin{split}
\tilde{x}&=-\sgn{Q_5} \times\\
         &\phantom{=}\begin{pmatrix}k_1&k_1(n,s_2Q_5,Q_5)\\k_2/k_3(s_2Q_5,n,ns_2)&(2k_1-ns_2k_2/k_3)Q_5\end{pmatrix}.
\end{split}
\end{gather}
While this dual includes denominators, they are guaranteed to cancel so it is a valid black hole. Clearly if both $k_1$ and $k_2$ vanish the entire system vanishes, failing to preserve rank. However, $k_1$ and $k_2$ can vanish separately and still leave a rank 4 system. In summary we have
\begin{equation}
Q_5\neq0\land(k_1\neq0\lor ns_2k_2\neq0),
\end{equation}
which is equal to its logical conjunction with \eqref{eq:senrank4cond} and is of the same form of \eqref{eq:duallargecanrank4cond} provided one remembers $s_1\propto k_1$. Further, using \eqref{eq:gcdrelns} we find
\begin{gather}
\begin{array}{r@{\ }c@{\ }c@{\ }c@{\ }l}
d_2(\tilde{x})         & = & d_2(x)  & = & \gcd(2n,Q_5),\\
d'_2(\tilde{x})        & = & d'_2(x) & = & \gcd(n,Q_5),\\
d_4(\tilde{x})         & = & d_4(x)  & = & Q_5{}^2|s_1{}^2-4ns_2|\\
                       &   &         & = & J^2-4nQ_1Q_5,\\
d_5(\tilde{x})         & = & d_5(x)  & = & 2|Q_5|\gcd(n,s_1Q_5,s_2Q_5)\\
                       &   &         & = & 2\gcd(nQ_5,JQ_5,Q_1Q_5),\\
r(\tilde{P},\tilde{Q}) & = & r(P,Q)  & = & |Q_5|,
\end{array}
\shortintertext{and}
\begin{array}{r@{\ =\ }l@{\!\!\!\!\!}c@{\ }l}
d_1(x)&1                     &d_1(\tilde{x})&=\gcd(k_1,k_2)\\
d_3(x)&|Q_5|\gcd(k_3,l)      &d_3(\tilde{x})&=k_0|Q_5|\\
      &\gcd(2nQ_5,2Q_1Q_5,J).&              &
\end{array}
\end{gather}
As expected \eqref{eq:d_icondition} is satisfied.

\subsection{Examples}
\label{sec:examples}

When considering examples we have three cases as in \autoref{sec:largeblacks}, namely $s_1=0\,\land\,ns_2\neq0$, $s_1\neq0\,\land\, ns_2=0$, and $ns_1s_2\neq0$. Examples satisfying these conditions and constraints \eqref{eq:senconditions} are listed in \autoref{tab:senexamples}. These are discussed in more detail in \hyperref[sec:moresenexamples]{appendix B}.
\begin{table*}
\caption[Large black hole F-duality in the Sen basis]{Examples of F-duality in the Sen basis. Parameters $k_0, k_1$ and $k_2$ are fixed by \eqref{eq:senconditions} up to a sign, and $k_3$ is given by \eqref{eq:sendef}. By inspection of \eqref{eq:sengeneralfdual}, $d_1(\tilde{x})$ is seen to be given by $\gcd(k_1,k_2)$ and for all the examples tabulated $d_1(\tilde{x})=1$. Note that we still require $k_0\neq0$ in all cases.}\label{tab:senexamples}
\begin{ruledtabular}
\begin{tabular}{cc*{8}{M{c}}c}
& Case  & s_1   & s_2     & n       & \sgn\Delta & k_0    & k_1      & k_2           & k_3                     & \\
\hline
& 1.1   & 0     & \pm n   & \neq0   & \pm        & 2|n|   & 0        & 1             & 2|n|                    & \\
& 1.2   & 0     & \neq0   & \pm s_2 & \pm        & 2|s_2| & 0        & 1             & 2|s_2|                  & \\
\hline
& 2.1   & \neq0 & ~~~0    & \neq0   & -          & |s_1|  & \sgn s_1 & 2|n/s_1|      & 2|n|                    & \\
& 2.2   & \neq0 & \neq0   & ~~~0    & -          & |s_1|  & \sgn s_1 & 2|s_2Q_5/s_1| & 2|s_2Q_5|               & \\
& 2.3   & \neq0 & ~~~0    & ~~~0    & -          & |s_1|  & \sgn s_1 & 1             & |s_1|                   & \\
\hline
& 3.1.1 & 2p    & p\pm1   & p\mp1   & -          & 2      & p        & k_3/2         & 2\gcd(p\mp1,(p\pm1)Q_5) & \\
& 3.1.2 & 2p    & p^2\pm1 & 1       & \pm        & 2      & p        & 1             & 2                       & \\
& 3.1.3 & 2p    & 1       & p^2\pm1 & \pm        & 2      & p        & k_3/2         & 2\gcd(p^2\pm1,Q_5)      & \\
& 3.2   & 2p+1  & p       & p+1     & -          & 1      & 2p+1     & k_3           & 2\gcd(p+1,pQ_5)         &
\end{tabular}
\end{ruledtabular}
\end{table*}

\section{\texorpdfstring{The $STU$ model}{STU model}}
\label{sec:STUmodel}

\subsection{\texorpdfstring{Jordan and FTS identities for $\mathfrak{J}=\mathds{Z}\oplus\mathds{Z}\oplus\mathds{Z}$}{Jordan and FTS identities for J=Z+Z+Z}}

In this section, we focus on the $STU$ model \cite{Duff:1995sm} which corresponds to $\mathfrak{J}=\mathds{Z}\oplus\mathds{Z}\oplus\mathds{Z}$ and for which the U-duality is $SL(2,\mathds{Z})\times SL(2,\mathds{Z}) \times SL(2,\mathds{Z})$.

\begin{enumerate}
\item For diagonal Jordan algebra elements $A=(A_1, A_2, A_3)$ and $B=(B_1, B_2, B_3)$ we have,
\begin{equation}
\begin{split}
A\circ B&=(A_1B_1,A_2B_2,A_3B_3)\\
N(A)&=A_1A_2A_3\\
\Tr(A)&=A_1+A_2+A_3\\
S(A)&=A_1A_2+A_2A_3+A_3A_1\\
S(A,B)&=\phantom{+}A_1(B_2+B_3)\\
      &\phantom{=\,}+A_2(B_3+B_1)\\
      &\phantom{=\,}+A_3(B_1+B_2)\\
\Tr(A,B)&=A_1B_1+A_2B_2+A_3B_3\\
A^\sharp &=(A_2A_3,A_3A_1,A_1A_2)\\
A\times B&=\begin{pmatrix}A_2B_3+A_3B_2\\A_3B_1+A_1B_3\\A_1B_2+A_2B_1\end{pmatrix}\\
N(A,B,C)&=\phantom{+}\tfrac{1}{6}(A_1(B_2 C_3+B_3 C_2)\\
        &\phantom{=\tfrac{1}{6}\ }+A_2(B_3 C_1+B_1 C_3)\\
        &\phantom{=\tfrac{1}{6}\ }+A_3(B_1 C_2+B_2 C_1)).
\end{split}
\end{equation}
\item For the most general case of the triple system with diagonal Jordan algebra entries $A=(A_1, A_2, A_3)$ and $B=(B_1, B_2, B_3)$,
    \begin{equation}
    x=\begin{pmatrix}\alpha & (A_1,A_2,A_3) \\(B_1,B_2,B_3) & \beta\end{pmatrix},
    \end{equation}
    we have $\kappa(x)=\half(\alpha\beta-(A_1B_1+A_2B_2+A_3B_3))$ and the quartic form $\Delta$ becomes
    \begin{equation}\label{eq:z+z+zhyperdet}
    \begin{split}
    \Delta(x)&=-4[\kappa^2+\alpha A_1A_2A_3+\beta B_1B_2B_3\\
    &\phantom{=-4[}-A_1 A_2 B_1 B_2-A_3 A_1 B_3 B_1\\&\phantom{=-4[}-A_2 A_3 B_2 B_3]\\
    &=-(\alpha^2\beta^2+A_1^2 B_1^2+A_2^2 B_2^2+A_3^2 B_3^2)\\
    &\phantom{=\ }+2 (\alpha\beta(A_1 B_1+A_2 B_2+A_3 B_3)\\
    &\phantom{=+2\ (}+A_1 A_2 B_1 B_2+A_3 A_1 B_3 B_1\\&\phantom{=+2\ (}+A_2 A_3 B_2 B_3)\\
    &\phantom{=\ }-4 (\alpha A_1 A_2 A_3 + \beta B_1 B_2 B_3)\\
    &=-4[(\kappa+A_1 B_1)^2-(\alpha A_1-B_2 B_3)\\
    &\phantom{=-4[(\kappa+A_1 B_1)^2\ }\times(\beta B_1 -A_2 A_3)]\\
    &=-4[(\kappa+A_2 B_2)^2-(\alpha A_2-B_3 B_1)\\
    &\phantom{=-4[(\kappa+A_1 B_1)^2\ }\times(\beta B_2 -A_3 A_1)]\\
    &=-4[(\kappa+A_3 B_3)^2-(\alpha A_3-B_1 B_2)\\
    &\phantom{=-4[(\kappa+A_1 B_1)^2\ }\times(\beta B_3 -A_1 A_2)],
    \end{split}
    \end{equation}
    where the all five forms exemplify triality. Finally, we have
    \begin{equation}
    \begin{split}
    T_\alpha&=-2(\alpha\kappa+B_1B_2B_3)\\
    T_\beta&=\phantom{-}2(\beta\kappa+2A_1A_2A_3)\\
    T_{A_1}&=-2(\beta B_2B_3-(A_2B_2+A_3B_3+\kappa)A_1)\\
    T_{A_2}&=-2(\beta B_3B_1-(A_3B_3+A_1B_1+\kappa)A_2)\\
    T_{A_3}&=-2(\beta B_1B_2-(A_1B_1+A_2B_2+\kappa)A_3)\\
    T_{B_1}&=\phantom{-}2(\alpha A_2A_3-(A_2B_2+A_3B_3+\kappa)B_1)\\
    T_{B_2}&=\phantom{-}2(\alpha A_3A_1-(A_3B_3+A_1B_1+\kappa)B_2)\\
    T_{B_3}&=\phantom{-}2(\alpha A_1A_2-(A_1B_1+A_2B_2+\kappa)B_3).
    \end{split}
    \end{equation}
\end{enumerate}
As well as describing  the 8 charges of the $STU$ model in full generality the above expressions also cover a generic FTS in diagonal reduced form.

The $STU$ model describes $\mathcal{N}=2$ supergravity coupled to three vector multiplets. Consequently, there are four electric charges $q$ and four magnetic charges $p$

\begin{equation}\label{eq:General8ParamStateInFTSForm}
x=\begin{pmatrix}-q_0 & (p^1,p^2,p^3) \\ (q_1,q_2,q_3) & p^0\end{pmatrix}.
\end{equation}
See \autoref{tab:Dictionaries} for a summary of the charges we assign to the FTS.
\begin{table*}
\caption[FTS dictionaries]{Assignments of values to a generic Freudenthal triple system.}\label{tab:Dictionaries}
\begin{ruledtabular}
\begin{tabular}{cc*{8}{M{c}}c}
& Basis      & \alpha          & \beta       & A_1             & A_2             & A_3             & B_1 & B_2 & B_3 & \\
\hline
& Canonical  & \alpha          & \alpha j    & \alpha k        & \alpha kl       & \alpha klm      & 0   & 0   & 0   & \\
& Projective &  1              & j           & 1               & 1               & k               & 0   & 0   & 0   & \\
& Sen        & -1\phantom{-}   & J=s_1Q_5    & n               & Q_1=s_2Q_5      & Q_5             & 0   & 0   & 0   & \\
& $STU$      & -q_0\phantom{-} & p^0         & p^1             & p^2             & p^3             & q_1 & q_2 & q_3 & \\
& Cayley     & -a_7\phantom{-} & a_0         & -a_1\phantom{-} & -a_2\phantom{-} & -a_4\phantom{-} & a_6 & a_5 & a_3 &
\end{tabular}
\end{ruledtabular}
\end{table*}
In this case,
\begin{gather}
\begin{split}
N(A)&=p^1p^2p^3,\\N(B)&=q_1q_2q_3,
\end{split}
\shortintertext{and}
\begin{split}
{A}^{\sharp }(P)&=(p^2p^3, p^1p^3, p^1p^2),\\ {B}^{\sharp }(Q)&=(q_2q_3, q_1q_3, q_1q_2),
\end{split}
\end{gather}
and $\Delta(x)$ of \eqref{eq:z+z+zhyperdet} becomes \cite{Behrndt:1996hu}
\begin{equation}\label{eq:Cayleyfreud}
\begin{split}
\Delta(x) &=-(p\cdot q)^2+4[(p^1q_1)(p^2q_2)+(p^1q_1)(p^3q_3)\\
          &\phantom{=\ }+(p^3q_3)(p^2q_2)- p^0 q_1 q_2 q_3 + q_0 p^1 p^2 p^3].
\end{split}
\end{equation}
Or using the transformation between $P,Q$ and $p,q$:
\begin{equation}\label{eq:pqtoPQdic}
\begin{bmatrix}
p^0 \\ p^1 \\ p^2 \\ p^3 \\ q_0 \\ q_1 \\ q_2 \\ q_3
\end{bmatrix}= \frac{1}{\sqrt{2}}
\begin{bmatrix*}[r]
P^0-P^2 \\ Q_0+Q_2 \\ P^3-P^1 \\ -P^3-P^1 \\ Q_0-Q_2 \\ -P^0-P^2 \\ Q_3-Q_1 \\ -Q_3-Q_1
\end{bmatrix*},
\end{equation}
under which we obtain the relations
\begin{gather}
\begin{array}{c@{\ =\ }c@{\ }c@{\ }c}
P^2      & 2(p^2p^3 & - & p^0q_1),   \\
P\cdot Q & p\cdot q & - & 2 p^1 q_1, \\
Q^2      & 2(p^1q_0 & + & q_2q_3).
\end{array}\label{eq:PQtoSen}
\shortintertext{then  we find}
\Delta(P,Q)=P^2 Q^2 -(P\cdot Q)^2,
\end{gather}
which is manifestly invariant under $SL(2) \times SO(2,2)$.

These eight charges may be usefully rewritten in the \emph{Cayley basis} as a $2 \times 2 \times 2$ hypermatrix $a_{ABC}$ \cite{Duff:1995sm}. In the black hole-qubit correspondence \cite{Duff:2006uz,Duff:2006ue,Kallosh:2006zs,Levay:2006kf,Levay:2006pt,Duff:2007wa,Bellucci:2007zi,Levay:2007nm} the $a_{ABC}$ are interpreted as the state vector coefficients of a three qubit system (Alice, Bob and Charlie).  Intriguingly, the FTS \autoref{tab:FTSrank} also provides the classification of different kinds of three-qubit entanglement \cite{Borsten:2008yb,Levay:2009}.

Performing a binary to decimal $a_0,\dotsc,a_7$ conversion on the indices of $a$, we have
\begin{gather}
\begin{array}{c@{\ \big(}*{7}{c@{,\ }}c@{\big)}c@{}c}\label{eq:charges7}
  & p^0   & p^1    & p^2    & p^3    & q_0   & q_1   & q_2   & q_3   & \\[3pt]
= & a_{0} & -a_{1} & -a_{2} & -a_{4} & a_{7} & a_{6} & a_{5} & a_{3} & ,
\end{array}
\end{gather}
under which
\begin{equation}
x=\begin{pmatrix}-a_{7}&-(a_{1},a_{2},a_{4})\\(a_{6},a_{5},a_{3})&a_{0}\end{pmatrix}.
\end{equation}
One finds that the quartic norm $\Delta(x)$ is related to Cayley's hyperdeterminant by
\begin{equation}\label{eq:q}
\Delta(x)=\det\gamma^{A}=\det\gamma^{B}=\det\gamma^{C}=:-\Det a,
\end{equation}
where, following \cite{Duff:2006ev,Toumazet:2006,Borsten:2008wd} we have defined the three matrices $\gamma^A,\gamma^B$, and $\gamma^C$
\begin{gather}
\begin{split}\label{eq:ABCgammas}
(\gamma^{A})_{A_{1}A_{2}}&=\varepsilon^{B_{1}B_{2}}\varepsilon^{C_{1}C_{2}}a_{A_{1}B_{1}C_{1}}a_{A_{2}B_{2}C_{2}}, \\
(\gamma^{B})_{B_{1}B_{2}}&=\varepsilon^{C_{1}C_{2}}\varepsilon^{A_{1}A_{2}}a_{A_{1}B_{1}C_{1}}a_{A_{2}B_{2}C_{2}}, \\
(\gamma^{C})_{C_{1}C_{2}}&=\varepsilon^{A_{1}A_{2}}\varepsilon^{B_{1}B_{2}}a_{A_{1}B_{1}C_{1}}a_{A_{2}B_{2}C_{2}}.
\end{split}
\end{gather}
transforming respectively as $\rep{(3,1,1), (1,3,1), (1,1,3)}$ under $SL(2) \times SL(2) \times SL(2)$. Explicitly,
\begin{widetext}
\begin{gather}
\begin{split}
\gamma^{A}&=
\begin{pmatrix}
2(a_{0}a_{3}-a_{1}a_{2}) &  a_{0}a_{7}-a_{1}a_{6}+a_{4}a_{3}-a_{5}a_{2}\\
a_{0}a_{7}-a_{1}a_{6}+a_{4}a_{3}-a_{5}a_{2}  & 2(a_{4}a_{7}-a_{5}a_{6})
\end{pmatrix}, \\
\gamma^{B}&=
\begin{pmatrix}2(a_{0}a_{5}-a_{4}a_{1}) & a_{0}a_{7}-a_{4}a_{3}+a_{2}a_{5}-a_{6}a_{1}\\
a_{0}a_{7}-a_{4}a_{3}+a_{2}a_{5}-a_{6}a_{1} & 2(a_{2}a_{7}-a_{6}a_{3})
\end{pmatrix}, \\
\gamma^{C}&=
\begin{pmatrix}
2(a_{0}a_{6}-a_{2}a_{4}) &  a_{0}a_{7}-a_{2}a_{5}+a_{1}a_{6}-a_{3}a_{4}\\
a_{0}a_{7}-a_{2}a_{5}+a_{1}a_{6}-a_{3}a_{4} & 2(a_{1}a_{7}-a_{3}a_{5})
\end{pmatrix},
\end{split}
\end{gather}
\end{widetext}
\begin{equation}\label{eq:CayleyHyperdeterminant}
\begin{split}
&\Det a \\ :=&-\half~\varepsilon^{A_1A_2}\varepsilon^{B_1B_2}\varepsilon^{A_3A_4}\varepsilon^{B_3B_4}\varepsilon^{C_1C_4}\varepsilon^{C_2C_3} \\
&\ \ \ \times a_{A_1B_1C_1}a_{A_2B_2C_2}a_{A_3B_3C_3}a_{A_4B_4C_4}\\
=&\phantom{-\ }a_{000}^2 a_{111}^2 + a_{001}^2 a_{110}^2+a_{010}^2 a_{101}^2 + a_{100}^2 a_{011}^2 \\
&-2\,(\phantom{+\,}a_{000}a_{001}a_{110}a_{111}+a_{000}a_{010}a_{101}a_{111} \\
&\phantom{2\,(\ \ \ }+a_{000}a_{100}a_{011}a_{111}+a_{001}a_{010}a_{101}a_{110}\\
&\phantom{2\,(\ \ \ }+a_{001}a_{100}a_{011}a_{110}+a_{010}a_{100}a_{011}a_{101})\\ &+4\,(a_{000}a_{011}a_{101}a_{110}+a_{001}a_{010}a_{100}a_{111})\\
=&\phantom{-\ }a_{0}^2 a_{7}^2 + a_{1}^2 a_{6}^2+  a_{2}^2 a_{5}^2 + a_{3}^2 a_{4}^2 \\
&-2\,(\phantom{-\ }a_{0}a_{1}a_{6}a_{7} +a_{0}a_{2} a_{5}a_{7} +a_{0}a_{4}a_{3}a_{7}\\
&\phantom{-2\,(\ }+a_{1}a_{2}a_{5}a_{6} +a_{1}a_{3}  a_{4}a_{6} +a_{2}a_{3}a_{4}a_{5}) \\
&+4\,(a_{0}a_{3}a_{5}a_{6}+ a_{1}a_{2}a_{4}a_{7}),
\end{split}
\end{equation}
$T_{ABC}$ takes one of three equivalent forms
\begin{equation}
\begin{split}
T_{A_3B_1C_1}=-\varepsilon^{A_1A_2}a_{A_1B_1C_1}(\gamma^{A})_{A_{2}A_{3}}\\
T_{A_1B_3C_1}=-\varepsilon^{B_1B_2}a_{A_1B_1C_1}(\gamma^{B})_{B_{2}B_{3}}\\
T_{A_1B_1C_3}=-\varepsilon^{C_1C_2}a_{A_1B_1C_1}(\gamma^{C})_{C_{2}C_{3}}.
\end{split}
\end{equation}
Explicitly,
\begin{equation}
\begin{split}
T_0&=a_0 \left(\phantom{-}a_3 a_4+a_2 a_5+a_1 a_6-a_0a_7\right)-2 a_1 a_2 a_4 \\
T_1&=a_1 \left(          -a_3 a_4-a_2 a_5+a_1a_6-a_0 a_7\right)+2 a_0 a_3 a_5 \\
T_2&=a_2 \left(          -a_3 a_4+a_2a_5-a_1 a_6-a_0 a_7\right)+2 a_0 a_3 a_6 \\
T_3&=a_3 \left(          -a_3a_4+a_2 a_5+a_1 a_6+a_0 a_7\right)-2 a_1 a_2 a_7 \\
T_4&=a_4 \left(\phantom{-}a_3a_4-a_2 a_5-a_1 a_6-a_0 a_7\right)+2 a_0 a_5 a_6 \\
T_5&=a_5 \left(\phantom{-}a_3 a_4-a_2a_5+a_1 a_6+a_0 a_7\right)-2 a_1 a_4 a_7 \\
T_6&=a_6 \left(\phantom{-}a_3 a_4+a_2 a_5-a_1a_6+a_0 a_7\right)-2 a_2 a_4 a_7 \\
T_7&=a_7 \left(          -a_3 a_4-a_2 a_5-a_1 a_6+a_0a_7\right) +2 a_3 a_5 a_6.
\end{split}
\end{equation}
Note
\begin{gather}
\gamma(T)=\det \gamma (a) \gamma(a),
\shortintertext{so}
\gamma(\tilde a)=\sgn(\Delta)  \gamma(a).\label{eq:gammaofduala}
\end{gather}
Defining
\begin{gather}
S^{A_1}{}_{A_2}=\varepsilon^{A_1A_3}\gamma_{A_3A_2}(\det\gamma)^{-1/2}
\shortintertext{we find}
\det S=\sgn(\Delta),
\end{gather}
and a Freudenthal duality cannot be undone by an $SL(2)$ duality in the non-BPS case $\Delta<0$.

\subsection{Examples}

\begin{description}
\item[Example 1] Choose
    \begin{equation}\label{eq:stughz}
    x=-\begin{pmatrix}a_{7}&(a_{1}, a_{2}, a_{4})\\(0, 0, 0)&0\end{pmatrix},
    \end{equation}
    in which case
    \begin{gather}
    \begin{split}
    \gamma^{A}&=2\begin{pmatrix}-a_{1}a_{2} &0\\0 & a_{4}a_{7}\end{pmatrix}, \\
    \gamma^{B}&=2\begin{pmatrix}-a_{4}a_{1} &0\\0 & a_{2}a_{7}\end{pmatrix}, \\
    \gamma^{C}&=2\begin{pmatrix}-a_{2}a_{4} &0\\ 0& a_{1}a_{7}\end{pmatrix},
    \end{split}
    \end{gather}
    and
    \begin{equation}
    \Delta(x)=-4a_{7}a_{1} a_{2} a_{4}=4q_0 p^1 p^2 p^3.
    \end{equation}

    The trilinear map yields
    \begin{equation}
    T(x)=-2\begin{pmatrix}0&(0, 0, 0)\\a_{7}(a_{2}a_{4},a_{1}a_{4}, a_{1}a_{2})&a_{1}a_{2}a_{4}\end{pmatrix},
    \end{equation}
    Setting $a_{7}=\pm a_{1}= n$ and $a_{2}=a_{4}=m$ so that
     \begin{equation}
    x=\begin{pmatrix}-n&-(\pm n, m, m)\\(0, 0, 0)&0\end{pmatrix},
    \end{equation}
      In this example $P^2=2m^2, P\cdot Q=0,Q^2=-2n^2$ and
    \begin{equation}
    \Delta(x)=- 4m^2n^2
    \end{equation}
    and the dual system $\tilde{x}$ is then given by
   \begin{equation}
    \tilde{x}=-\sgn(mn)\begin{pmatrix}0&
    (0, 0, 0)\\
    (m, \pm n, \pm n)&
    \pm m\end{pmatrix}.
    \end{equation}
  \item[Example 2] Choose
    \begin{equation}\label{eq:stuspecial1}
    x=\begin{pmatrix}-a_{7}&(-a_{1}, 0, 0)\\(a_6, 0, 0)&a_0\end{pmatrix},
    \end{equation}
    in which case
    \begin{gather}
    \begin{split}
    \gamma^{A}&=\begin{pmatrix}0& a_{0}a_{7}-a_{1}a_{6}   \\a_{0}a_{7}-a_{1}a_{6} & 0\end{pmatrix}, \\
    \gamma^{B}&=\begin{pmatrix}0& a_{0}a_{7}-a_{1}a_{6}   \\a_{0}a_{7}-a_{1}a_{6} & 0\end{pmatrix}, \\
    \gamma^{C}&=\begin{pmatrix}2a_{0}a_{6} & a_{0}a_{7}+a_{1}a_{6}\\ a_{0}a_{7}+a_{1}a_{6}& 2a_{1}a_{7}\end{pmatrix},
    \end{split}
    \end{gather}
    In this example $P^2=0, P\cdot Q=a_{0}a_{7} -a_{1} a_{6},Q^2=-a_1a_7$ and
    \begin{equation}
    \Delta(x)=-(a_{0}a_{7} -a_{1} a_{6})^2
    \end{equation}
    and the dual system $\tilde{x}$ is then given by
    \begin{equation}
    \tilde x=\begin{pmatrix}a_{7}&(-a_{1}, 0, 0)\\(-a_6, 0, 0)&a_0\end{pmatrix}.
    \end{equation}
\end{description}

\section{\texorpdfstring{The 5D Jordan dual}{The 5D Jordan dual}}
\label{sec:jdual}

\subsection{Definition}

Given a black string with charges $A$ or black hole with charges $B$, we define its Jordan dual by
\begin{align}
A^\star&={A^\sharp}{N(A)}^{-1/3},& B^\star&={B^\sharp}{N(B)}^{-1/3},
\end{align}
where we take the real root as implied by the notation. As described in \autoref{sec:J}, the Jordan algebra divides black holes and strings into four distinct ranks or orbits. J-duality is initially defined for large rank 3 strings for which both ${A^\sharp}$  and $N(A)$ are nonzero and large rank 3 holes for which both ${B^\sharp}$  and $N(B)$ are nonzero.
Small black holes and strings are discussed in \autoref{sec:Smith}, we also discuss an alternative definition of the Jordan dual in \hyperref[sec:altJordanDual]{appendix D}.

The invariance of $N(A)$ follows by noting that
\begin{equation}
\Tr(A^\sharp, A)=3N(A),
\end{equation}
where $A^\sharp$ obeys
\begin{equation}\label{eq:sharpofsharpofx}
(A^\sharp)^\sharp=N(A) A,
\end{equation}
and hence
\begin{equation}
N(A^\sharp)=N(A)^2.
\end{equation}
So
\begin{equation}
N(A^\star)=N(A^\sharp N(A)^{-1/3})=N(A).
\end{equation}
Moreover
\begin{equation}
A^{\star\star}=(A^\sharp N(A^\sharp)^{-1/3})^\sharp N(A^\star)^{-1/3}=A.
\end{equation}
Similar results hold for $B$.

In the case of a black holes and black string related by Jordan duality, the Dirac-Schwinger quantisation condition \eqref{eq:DSJ} is given by
\begin{equation}
\Tr(A^\star, A)=3N(A)^{2/3},
\end{equation}
which is also invariant.  Note the factor of 3.

As noted in \autoref{sec:Introduction}, for a valid dual $A^\star$, we require that $N(A)$ is a perfect cube. This is a necessary, but not sufficient condition because we further require that
\begin{equation}
d_3(A)=\left[\frac{d_2(A)}{d_1(A^\star)}\right]^3
=\left[\frac{d_2(A^\star)}{d_1(A^)}\right]^3=d_3(A^\star).
\end{equation}

In the 5D case the Smith diagonal form  of \eqref{eq:Smith} is unique in the sense that it is unambiguously determined by the U-duality invariants $d_1(A)$, $d_2(A)$ and $N(A)$.
\begin{quote}
\emph{Black holes related by a J-duality not conserving $d_1(A)$ provide examples of configurations with the same cubic norm and hence lowest order entropy that are not U-duality related.}
\end{quote}

The U-duality integral invariants $\Tr(X,Y)$ and $N(X,Y,Z)$ are not generally invariant under Jordan duality while $\Tr(A^\star,A)$ and $N(A)$,  and hence the lowest-order black hole entropy are. However, higher order corrections to the black hole entropy depend on some of the discrete U-duality invariants, to which we now turn.

\subsection{The action of J-duality on discrete U-duality invariants}
\label{sec:Jaction}

J-duality commutes with U-duality in the sense that $A^\star$ transforms contragredient to $A$. This follows from the property that a linear transformation $s$ belongs to the norm preserving group if and only if
\begin{equation}
s(A) \times s(B)= s'(A\times B),
\end{equation}
where $s'$ is given by
\begin{equation}
\Tr(s(A), s'(B))=\Tr(A, B)
\end{equation}
and always belongs to the norm preserving group if $s$ itself does \cite{Springer:2000}. This implies
\begin{equation}
(s(A))^\star= s'(A^\star).
\end{equation}
As we shall see in the following section, of the discrete invariants listed in \eqref{eq:DiscreteInvariants5}, only the cubic norm $d_3(A)$ is generically preserved under J-duality.

\subsection{Smith diagonal form and its dual}
\label{sec:Smith}

We have already seen in \autoref{sec:gcds} that we may write the most general black string charge configuration, up to U-duality, as
\begin{equation}\label{eq:can5Dblack}
A= k  (1, l, lm),
\end{equation}
where $k, l \geq 0$. In this case
\begin{equation}
A^\sharp= k^2l  (lm, m, 1),
\end{equation}
and
\begin{equation}
N(A)=k^3l^2m.
\end{equation}
So the Jordan dual black string is given by
\begin{equation}\label{eq:dualcan5Dblack}
A^\star = k ({l}/{m})^{1/3} (lm, m, 1).
\end{equation}
Hence, we require $k^3l=n^3|m|,\,n\in\mathds{N}$. The general $A$ and $A^\star$ related by J-duality are then
\begin{equation}
\begin{split}\label{eq:can5Dblack*}
A&= k  (1, l, lm)\\
A^\star&=n (lm, m, 1),
\end{split}
\end{equation}
with $\gcd$s
\begin{equation}
\begin{aligned}
d_1(A)&=k         & d_1(A^\star)&=n          \\
d_2(A)&=k^2l      & d_2(A^\star)&=n^2|m|   \\
d_3(A)&=k^3l^2|m| & d_3(A^\star)&=n^3m^2l.
\end{aligned}
\end{equation}
So $d_3(A)$ is conserved as expected and so is the product $d_1(A)d_2(A)$ but not $d_1(A)$ and $d_2(A)$ separately, except when $n=k$.

The similar form of $A$ and $A^\star$ when $n=k$ suggests they may be related. In fact they must be related by a U-duality because they have the same $d_1$, $d_2$ and $d_3$.

Note that $N^2$ is a perfect cube
\begin{equation}
N^2=(nklm)^3,
\end{equation}
which also implies that $N$ is a perfect cube, as can be deduced by considering its prime decomposition, consistent with the claim in \autoref{sec:Introduction}.

For large black holes there is an unambiguous J-dual stemming from the fact that both $A^\sharp$ and $N(A)$ are nonzero. For rank 2 we have $N(A)=0$ but $A^\sharp\neq0$ and we do not expect a J-dual to exist. For lower ranks both quantities vanish and since $A^\star$ is a vanishing quadratic quantity over the cube root of a vanishing cubic quantity, we might expect a finite result.  As in 4D, however, the result is not unique. Putting $m=l$ and then setting $l=0$ in \eqref{eq:can5Dblack} and \eqref{eq:can5Dblack*} yields
\begin{equation}
\begin{split}
A&= k (1, 0, 0)\\
A^\star&=k (0, 0, 1),
\end{split}
\end{equation}
which are both rank 1. But putting $l=0$ yields
\begin{equation}
\begin{split}
A&= k  (1, 0, 0)\\
A^\star&=0.
\end{split}
\end{equation}
So $A$ is the same but the dual is rank 0.  As in the Freudenthal case, therefore, this apparent lack of uniqueness favours continuing to restrict J-duality to large black holes/strings.

\section{Freudenthal/Jordan duality and the 4D/5D lift}
\label{sec:FJlift}

\subsection{Reduced element}

We recall that a black hole can be put into reduced form:
\begin{equation} \label{eq:reduced4d5d}
        x = \begin{pmatrix} \alpha  &  A \\
                            0 &   \beta \end{pmatrix}.
\end{equation}
We now show that for these five parameter black holes the lift of the Freudenthal dual is related to the Jordan dual. For the black hole in \eqref{eq:reduced4d5d} we have
\begin{equation}
\begin{split}
\Delta(x) &= -\alpha^2\beta^2 -4\alpha N(A), \\
T(x)      &=  \begin{pmatrix}-\alpha^2 \beta      & \alpha\beta A \\
                             2\alpha A^\sharp    & \alpha\beta^2 + 2N(A)
              \end{pmatrix}.
\end{split}
\end{equation}
We have the following $\cP(x)$ and $\cQ(x)$
\begin{equation}\label{eq:PQ}
\begin{split}
\cP(x) &= B^\sharp - \alpha A=  - \alpha A\\
\cQ(x) &= A^\sharp - \beta B =  A^\sharp,
\end{split}
\end{equation}
with the following norms
\begin{align}\label{eq:NPNQ}
N(\cP(x)) &= - \alpha^3 N(A), & N(\cQ(x)) &= N(A)^2,
\end{align}
and the following angular momenta
\begin{equation}
\begin{split}
\cJ_\alpha&= -\half T_\alpha=\phantom{-}\half\alpha^2\beta, \\
\cJ_\beta &= -\half T_\beta=-\half\alpha\beta^2-N(A).
\end{split}
\end{equation}

The Freudenthal dual of $x$ is given by
\begin{equation}
\begin{split}
\tilde{x}&=\begin{pmatrix}\tilde\alpha & \tilde{A} \\
                         \tilde{B} & \tilde\beta \end{pmatrix}\\
         &= \frac{1}{\abs{\Delta}^{1/2}}  \begin{pmatrix} -\alpha^2 \beta   & \alpha\beta A \\
                                                          2\alpha A^\sharp  & \alpha\beta^2 + 2N(A)\end{pmatrix}.
\end{split}
\end{equation}
Hence
\begin{align}
\tilde{A}^\sharp &= \frac{ \alpha^2\beta^2 A^\sharp }{|\Delta|}, & \tilde{B}^\sharp &= \frac{ 4 \alpha^2 N(A) A }{\abs{\Delta}}.
\end{align}
So we have the following $\cP(\tilde x)$ and $\cQ(\tilde x)$
\begin{gather}
\begin{split}
\cP(\tilde{x}) &= \tilde{B}^\sharp - \tilde\alpha \tilde{A} \\
               &= \frac{ 4 \alpha^2 N(A) A }{\abs{\Delta}} - \frac{-\alpha^2 \beta}{\abs{\Delta}^{1/2}} \cdot %
                \frac{ \alpha\beta A}{\abs{\Delta}^{1/2}} \\
               &=  \alpha \frac{(\alpha^2\beta^2 +  4 \alpha N(A)}{\abs{\Delta}}A \\
               &=  - \sgn(\Delta) \alpha A,
\end{split}
\shortintertext{and}
\begin{split}
\cQ(\tilde{x})  &= \tilde{A}^\sharp - \tilde\beta \tilde{B} \\
          &=  \frac{ \alpha^2\beta^2 A^\sharp }{|\Delta|} -  \frac{ \alpha\beta^2 + 2N(A)}{|\Delta|^{1/2}} \cdot %
            \frac{2\alpha A^\sharp}{\abs{\Delta}^{1/2}}\\
          &=  - \frac{(\alpha^2\beta^2 +  4 \alpha N(A)}{\abs{\Delta}}A^\sharp  \\
          &=  \sgn(\Delta) A^\sharp.
\end{split}
\end{gather}
Hence
\begin{align}
\cP(\tilde{x}) &= \sgn(\Delta) \cP(x), & \cQ(\tilde{x}) &= \sgn(\Delta) \cQ(x),
\end{align}
as expected from \eqref{eq:Ptilde} and \eqref{eq:Qtilde}. Similarly we find
\begin{align}
{\cJ_\alpha}(\tilde x)&= |\Delta|^{1/2}\alpha, & {\cJ_\beta}(\tilde x)&= |\Delta|^{1/2}\beta,
\end{align}
so that
\begin{equation}
\begin{split}
\Delta(\tilde x)&=  {|\Delta|}^{}{\cJ_\alpha}^{-2}\left(4\sgn(\Delta) N\left(\cP\right) - \abs{\Delta} \alpha^2 \right)\\
                &= \Delta(x).
\end{split}
\end{equation}

Now, if we take the Jordan duals of $\cP(x)$ and $\cQ(x)$, we have
\begin{align}
\cP^\star(x) &= \frac{\cP(x)^\sharp}{N(\cP(x))^{1/3}},     &    \cQ^\star(x) &= \frac{\cQ(x)^\sharp}{N(\cQ(x))^{1/3}}.
\end{align}
We can calculate $\cP^\sharp$ and $\cQ^\sharp$ from \eqref{eq:PQ}, for which we get $\cP^\sharp = \alpha^2A^\sharp$ and $\cQ^\sharp = N(A)A$, we already know $N(\cP)$ and $N(\cQ)$ from \eqref{eq:NPNQ}, so that we now have
\begin{equation}
\begin{aligned}
\cP^\star(x) &= \frac{\alpha^2 A^\sharp}{(-\alpha^3 N(A))^{1/3}} &\cQ^\star(x) &= \frac{N(A)A}{N(A)^{2/3}}\\
             &= -\frac{\alpha }{N(A)^{1/3}} A^\sharp,            &          &= N(A)^{1/3}A,
\end{aligned}
\end{equation}
with norms
\begin{align}\label{eq:}
N(\cP^\star) &= - \alpha^3 N(A) & N(\cQ^\star) &= N(A)^2.
\end{align}
Putting all this together, we find
\begin{equation}\label{eq:final}
\begin{split}
\hat{\cP}^\star(x) &= \hat{\cQ}(\tilde{x}),  \\
\hat{\cQ}^\star(x) &= \hat{\cP}(\tilde{x}),
\end{split}
\end{equation}
where the hat denotes an element with the unit norm;
\begin{align}
\hat{X} &= \frac{X}{N(x)^{1/3}}, & N(\hat{X})&=1.
\end{align}

Thus we have established
\begin{widetext}
\begin{equation}
\begin{CD}
\text{4D black hole $x$} @>\text{4D/5D lift}>> \text{5D black string $A\sim {\tilde B}^\star$}\\
@V{\text{Freudenthal dual}}VV @VV{\text{Jordan dual}}V\\
\text{dual 4D black hole $\tilde x$} @>>\text{4D/5D lift}>  \text{dual 5D black hole $\tilde B\sim A^\star$}
\end{CD}
\end{equation}
\end{widetext}

\subsection{Example}

To discuss J-duality and F-duality  simultaneously, we need $N(A)$ a perfect cube and
$\Delta(x)$ a perfect square. So we begin in canonical form (and assume $m$ positive for simplicity) with $j=0$, $k=p^2m$ and $l=q^3m$. So for $x$ we have
\begin{equation}
\begin{split}
x&=\phantom{p}\alpha\begin{pmatrix}1&p^2m(1,q^3m,q^3m^2)\\0&0\end{pmatrix},\\
\Delta&=-4p^6q^6\alpha^4m^6\\
\tilde{x}&=\alpha p\begin{pmatrix}0&(0,0,0)\\(q^3m^2,m,1)&p^2q^3m^3\end{pmatrix}.
\end{split}
\end{equation}
While for $A$ we have
\begin{equation}
\begin{split}
A&= \alpha p^2m  (1, q^3m,q^3m^2)\\
N(A)&=q^6p^6\alpha ^3m^6\\
A^\star&= q(\alpha p^2 m)  (q^3m^2, m, 1)=mqp {\tilde B}.
\end{split}
\end{equation}
So $d_1(\tilde x)=p\,d_1(x)\ \forall\ q$ in 4D but $d_1(A^\star)=q\,d_1(A)\ \forall\, p$ in 5D.

We find
\begin{align}
\begin{split}
\cP(A,B)&=-\alpha^2 p^2m  (1, q^3m,q^3m^2)\\
\cQ(A,B)&= \alpha^2 p^4m^3q^3  (q^3m^2, m, 1),
\end{split}
\shortintertext{and}
\begin{split}
\cP(\tilde A, \tilde B)&=\alpha^2 p^2m  (1, q^3m,q^3m^2)\\
\cQ(\tilde A, \tilde B)&=-\alpha^2 p^4m^3q^3  (q^3m^2, m, 1),
\end{split}
\shortintertext{and}
\begin{split}
\cP(A,B)^\star&=  -\alpha^2 p^2mq  (q^3m^2, m, 1)\\
\cQ(A,B)^\star&= \alpha^2 q^2p^4m^3  (1, q^3m,q^3m^2).
\end{split}
\end{align}

Hence \eqref{eq:final} is confirmed.

\section{Conclusions}
\label{sec:conclusions}

\subsection*{\texorpdfstring{$\mathcal{N}=8$:}{N=8}}

In the subcases where $d_1(x)$ is conserved, F-duality $x \to \tilde x$ preserves all the U-duality invariants \eqref{eq:DiscreteInvariants}. The degeneracy formula for the class of black holes considered in \cite{Sen:2008sp} depends explicitly on only $\Delta(x)$ and $d_5(x)$ and therefore the exact entropy in this case is F-dual invariant. The more general case remains an open question since we are not aware of a general U-duality invariant expression for dyon degeneracies.

In the projective case, this result is somewhat trivial because all black holes are U-duality related and so, in particular,  the F-dual $\tilde x$ is U-dual equivalent to $x$. The explicit U-duality is given in \hyperref[sec:FDualUDualUndo]{appendix C}.

In the non-projective case, this result seems non-trivial because we are not aware of any argument that would indicate that the F-dual $\tilde x$ is U-dual equivalent to $x$. For example the negative $j$ branch of case 2.1 of \autoref{tab:examples_table}:
\begin{equation}\label{eq:example}
\begin{split}
x        &=\alpha\begin{pmatrix}1 & (0,0,0) \\ (0,0,0) & j\end{pmatrix}, \\
\tilde{x}&=\alpha\begin{pmatrix}1 & (0,0,0) \\ (0,0,0) &-j\end{pmatrix}.
\end{split}
\end{equation}
Without a complete orbit classification the U-equivalence, or not, of F-dual black holes is a difficult question to answer in general. Even with a full orbit classification the invariance of the higher-order corrections to the entropy would remain unsettled as
we cannot be sure on which invariants they depend.
 Could there be black holes with the same precision entropy that are not U-duality related but are F-duality related?

In the subcases where  $d_1(x)$ is not conserved, we can be absolutely sure that the F-dual $\tilde x$ is not U-dual equivalent to $x$. In this case, however, we do not know whether F-duality leaves higher order corrections invariant because all the treatments of higher-order corrections we are aware of are restricted to  $d_1(x)=1$.
\begin{table}
\caption{Are F or J duals related by U-duality?}\label{tab:U}
\begin{ruledtabular}
\begin{tabular}{*{6}{c}}
& Duality & $d_1$ conserved ? &                & U-dual ? & \\
\hline
& \multirow{2}{*}{F-dual} & \multirow{2}{*}{Yes} & Projective     & Yes      & \\
&                         &                      & Non-projective & ?        & \\
& F-dual                  & No                   &                & No       & \\
& J-dual                  & Yes                  &                & Yes      & \\
& J-dual                  & No                   &                & No       &
\end{tabular}
\end{ruledtabular}
\end{table}

These 4D conclusions, and the simpler 5D ones, are summarised in \autoref{tab:U}.

\subsection*{\texorpdfstring{$\mathcal{N}=4$, heterotic:}{N=4, heterotic}}

F-duality $x \to \tilde x$ leaves invariant $\Delta$ and (up to a sign) $P^2$, $Q^2$ and $P\cdot Q$. Moreover, the discrete torsion $r(P,Q)$ is invariant. This result seems non-trivial because we are not aware of any argument that would indicate that the F-dual $\tilde x$ is T-dual equivalent to $x$. In the cases where $P^2$, $Q^2$ and $P\cdot Q$ flip sign, we can be absolutely sure that the F-dual $\tilde x$ is not T-dual equivalent to $x$. This corresponds specifically to  non-BPS black holes and, hence, the conjectured counting formula for all 1/4-BPS dyons is not applicable. However, it is perhaps encouraging that torsion is left invariant as it plays a central role in the current $\mathcal{N}=4$ dyon degeneracy calculations \cite{Banerjee:2008pu}.

In the subcases where $d_1(x)$ is not conserved, we can be absolutely sure that the F-dual $\tilde x$ is not U-dual equivalent to $x$. In this case, however, we do not know whether F-duality leaves higher order corrections invariant because all the treatments of higher-order corrections we are aware of are restricted to  $d_1(x)=1$. This restriction is typically imposed to avoid complications arising from the possibility that dyons with $d_1(x)>1$ may decay into single particle states. The consequences of this phenomenon for F-dual black holes remains an open question.

\subsection*{\texorpdfstring{$\mathcal{N}=2$, magic:}{N=2, magic}}
The magic $\mathcal{N}=2$ black holes may require a separate analysis since the diagonally reduced form, central to our present treatment, is not necessarily applicable in these instances. In particular, for the octonionic $\mathcal{N}=2$ example (as opposed to the \emph{split}-octonionic $\mathcal{N}=8$ case) it is well know that there are integral Jordan algebra elements that cannot be diagonalised \cite{Elkies:1996, Gross:1996, Krutelevich:2002}.

\subsection*{Further work}

For the time being the microscopic stringy interpretation of F-duality remains unclear. In part, this is due to the F-duality action only being defined on the black hole charges and not the component fields of the lowest order action. Having specified the necessary and sufficient conditions \eqref{eq:threeconditions} required for a well defined F-dual charge vector, one might ask how this space of black holes is mathematically characterised and whether it has a broader significance.

\begin{acknowledgments}

This work was supported in part by the STFC under rolling grant ST/G000743/1. We are grateful to Hajar Ebrahim and Dan Waldram for many illuminating discussions on black holes and to Sergio Ferrara for correspondence on the conditions under which  $\Delta$ is a perfect square.

\end{acknowledgments}

\appendix

\section{More  examples in canonical basis}
\label{sec:more}

Here we provide further discussion of subcases 2.3 and 3.1 of \autoref{sec:largeblacks}.

\begin{description}
\item[Subcase 2.3:] To simplify our considerations we choose to instead decompose $j$ in terms of $k,l$ and $\alpha$:
    \begin{gather}
    j=\pm2^{p_1}k^{p_2}l^{p_3}\alpha^{p_4},\quad p_i\in\mathds{N}_0,\label{eq:case2jansatz}
    \shortintertext{under which}
    \begin{gathered}
    n_2=\sgn j2^{1-p_1}k^{2-p_2}l^{1-p_3}\alpha^{1-p_4},\\
    p_1,p_3,p_4\in\{0,1\},\ p_2\in\{0,1,2\}.
    \end{gathered}
    \end{gather}
    This treatment therefore encompasses the $2^3\cdot3=24$ most obvious cases, but it leaves
    \begin{equation}
    d_1(\tilde{x})=\begin{cases}\alpha&p_4=0\\\gcd(\alpha,2^{1-p_1}k^{2-p_2}l^{1-p_3})&\text{else.}\end{cases}
    \end{equation}
    To further evaluate the $\gcd$ we need to make an ansatz for $\alpha$:
    \begin{gather}
    \alpha=2^{q_1}k^{q_2}l^{q_3}q_4,\quad q_{1,2,3}\in\mathds{N}_0,q_4\in\mathds{Z}\setminus\{0\},
    \shortintertext{under which}
    \begin{split}
    j&=\pm2^{p_1+q_1p_4}k^{p_2+q_2p_4}l^{p_3+q_3p_4}q_4{}^{p_4}\\
    n_2&=\sgn j2^{1-p_1+q_1(1-p_4)}k^{2-p_2+q_2(1-p_4)}\\
       &\phantom{=\ }\times l^{1-p_3+q_3(1-p_4)}q_4{}^{1-p_4}\\
    d_1(\tilde{x})&=2^{\min(q_1,1-p_1)}k^{\min(q_2,2-p_2)}l^{\min(q_3,1-p_3)}.
    \end{split}
    \end{gather}

\item[Subcase 3.1:] Starting with $j=2p$ we may postulate $p=rl$ as in subcase 3.1.2, but this time leave $k$ arbitrary to obtain
    \begin{equation}\label{eq:diophantineexample}
    \begin{split}
    n_0&=\pm2l[r^2+k^3m]^{1/2}=2lq,\ q\in\mathds{N},\\
    n_1&=\alpha r/q,\\
    n_2&=\alpha k^2/q,\\
    \sgn\Delta&=-\sgn(r^2+k^3m).
    \end{split}
    \end{equation}
    In general, suitable values of $\alpha$ must be chosen to enforce $n_1,n_2\in\mathds{Z}$. In \autoref{tab:diophantineexamples1} we list all charge vectors satisfying the ansatz $j=2rl\land |r^2+k^3m|=q^2$ where we have the absolute values of all parameters to be $\leq5$. This restriction is motivated by space constraints rather than any difficulty in finding more examples. Beyond these particular cases we may simply search for solutions to $|j^2+4k^3 l^2 m|=p^2, p\neq0$. In \autoref{tab:diophantineexamples2} we list all examples for which the absolute parameter values are $\leq3$.

\begin{table}
\caption[Large canonical black hole F-duals with restricted Diophantine parameters]{Charge vector parameters for \eqref{eq:diophantineexample}, restricting to small parameter values, where $s\in\mathds{N}$. This is an exhaustive list for $|r|,k,|m|,q$ values $\leq5$}\label{tab:diophantineexamples1}
\begin{ruledtabular}
\begin{tabular}{c*{10}{M{c}}c}
& \alpha & |r| & k & m    & q   & \sgn\Delta & n_0 & |n_1| & n_2 & d_1(\tilde{x}) & \\
\hline
& 2s     & 1   & 1 & -5~~ & 2   & +          & 4l  & s     & s   & s              & \\
&        & 1   & 1 & -2~~ & 1   & +          & 2l  & 1     & 1   & 1              & \\
& 2s     & 1   & 1 & 3    & 2   & -          & 4l  & s     & s   & s              & \\
& 3s     & 1   & 2 & 1    & 3   & -          & 6l  & s     & 4s  & s              & \\
& 5s     & 1   & 2 & 3    & 5   & -          & 10l & s     & 4s  & s              & \\
&        & 2   & 1 & -5~~ & 1   & +          & 2l  & 2     & 1   & 1              & \\
&        & 2   & 1 & -3~~ & 1   & -          & 2l  & 2     & 1   & 1              & \\
& 3s     & 2   & 1 & 5    & 3   & -          & 6l  & 2s    & s   & s              & \\
&        & 2   & 2 & -1~~ & 2   & +          & 4l  & 1     & 2   & 1              & \\
& 2s     & 3   & 1 & -5~~ & 2   & -          & 4l  & 3s    & s   & s              & \\
&        & 3   & 2 & -1~~ & 1   & -          & 2l  & 3     & 4   & 1              & \\
& 5s     & 3   & 2 & 2    & 5   & -          & 10l & 3s    & 4s  & s              & \\
&        & 4   & 2 & -4~~ & 4   & +          & 8l  & 1     & 1   & 1              & \\
&        & 5   & 2 & -3~~ & 1   & -          & 2l  & 5     & 4   & 1              & \\
& 3s     & 5   & 2 & -2~~ & 3   & -          & 6l  & 5s    & 4s  & s              &
\end{tabular}
\end{ruledtabular}
\end{table}
\begin{table}
\caption[Large canonical black hole F-duals with general Diophantine parameters]{Charge vector parameters resulting in integral dual charge vectors. This is an exhaustive list for $|p|,k,|m|,|q|$ values $\leq3$}\label{tab:diophantineexamples2}
\begin{ruledtabular}
\begin{tabular}{c*{10}{M{c}}c}
& \alpha & |j| & k & l & m    & \sgn\Delta & n_0 & |n_1| & n_2 & d_1(\tilde{x}) &\\
\hline
& 1      & 2   & 1 & 1 & -2~~ & +          & 2   & 1     & 1   & 1              &\\
& 1      & 3   & 1 & 1 & -2~~ & -          & 1   & 3     & 2   & 1              &\\
& 2      & 2   & 1 & 1 & -2~~ & +          & 2   & 2     & 2   & 2              &\\
& 2      & 2   & 1 & 1 & 3    & -          & 4   & 1     & 1   & 1              &\\
& 2      & 3   & 1 & 1 & -2~~ & -          & 1   & 6     & 4   & 2              &\\
& 3      & 1   & 1 & 1 & 2    & -          & 3   & 1     & 2   & 1              &\\
& 3      & 2   & 1 & 1 & -2~~ & +          & 2   & 3     & 3   & 3              &\\
& 3      & 2   & 1 & 2 & 2    & -          & 6   & 1     & 2   & 1              &\\
& 3      & 2   & 2 & 1 & 1    & -          & 6   & 1     & 4   & 1              &\\
& 3      & 3   & 1 & 1 & -2~~ & -          & 1   & 9     & 6   & 3              &\\
& 3      & 3   & 1 & 3 & 2    & -          & 9   & 1     & 2   & 1              &
\end{tabular}
\end{ruledtabular}
\end{table}

\end{description}

\section{Examples in Sen basis}
\label{sec:moresenexamples}

\subsubsection*{\texorpdfstring{Case 1: $s_1=0\land ns_2\neq0$}{Case 1: s1=0 and ns2 nonzero}}

We initially obtain
\begin{equation}
\begin{split}
k_0&=2|ns_2|^{1/2},\\
k_1&=0,\\
k_2&=\gcd(n,s_2Q_5)|ns_2|^{-1/2},\\
k_3&=2\gcd(n,s_2Q_5).
\end{split}
\end{equation}
To force $k_0\in\mathds{Z}$ we require $ns_2=\pm p^2, p\in\mathds{Z}$. Clearly then, we are not able to obtain an exhaustive decomposition as we did in the $j=0$ case of \autoref{sec:largeblacks}.

To progress we postulate that $n$ and $s_2$ are proportional to each other, with a perfect square constant of proportionality:
\begin{description}
\item[Subcase 1.1: $s_2=\pm p^2n^{2q+1}$.]
    \begin{equation}
    \begin{split}
    k_0&=2|pn^{q+1}|,\\
    k_2&=|pn^{q}|^{-1},\\
    k_3&=2|n|.
    \end{split}
    \end{equation}
    Then $p=\pm1, q=0$ so that $s_2=\pm n,k_0=2|n|,k_2=1$ as in \autoref{tab:senexamples}.

\item[Subcase 1.2: $n=\pm p^2s_2{}^{2q+1}$.]
    \begin{equation}
    \begin{split}
    k_0&=2|ps_2{}^{q+1}|,\\
    k_2&=|ps_2{}^{q}|^{-1}\gcd(p^2s_2{}^{2q},Q_5),\\
    k_3&=2|s_2|\gcd(p^2s_2{}^{2q},Q_5).
    \end{split}
    \end{equation}
    By again choosing $p=\pm1,q=0$ we obtain $n=\pm s_2,k_0=k_3=2|s_2|,k_2=1$. In contrast to case 1.1 we are not however forced to choose these $p,q$ values since an ansatz for $Q_5$ can satisfy the $k_2$ integer requirement:
    \begin{gather}
    Q_5=p^{r_1}s_2{}^{r_2}s,\quad r_i\in\mathds{N}_0,s\in\mathds{Z}\setminus\{0\},
    \shortintertext{under which}
    \begin{split}
    d_1(\tilde{x})\equiv k_2&=|p^{t_1-1}s_2{}^{t_2-q}|\times\\
    &\phantom{=\ }\gcd(p^{2-t_1}s_2{}^{2q-t_2},p^{r_1-t_1}s_2{}^{r_2-t_2}s),\\
    k_3&=2|p^{t_1}s_2{}^{t_2+1}|\times\\
    &\phantom{=\ }\gcd(p^{2-t_1}s_2{}^{2q-t_2},p^{r_1-t_1}s_2{}^{r_2-t_2}s),
    \end{split}
    \shortintertext{where we have defined}
    t_1:=\min(2,r_1) \text{\ \ and\ \ } t_2:=\min(2q,r_2).
    \end{gather}
    Insisting on $t_1,t_2$ (and hence $r_1,r_2$) satisfying $k_2,k_3\in\mathds{Z}$ results in
    \begin{widetext}
    \begin{equation}
    \begin{split}
    k_2&=\begin{cases}|ps_2{}^q|                                      & 2\leq r_1\phantom{\leq2\;} ~\land~ 2q\leq r_2 \\
                      |ps_2{}^{r_2-q}|\gcd(s_2{}^{2q-r_2},p^{r_1-2}s) & 2< r_1\phantom{\leq2\;} ~\land~ \phantom{2}q\leq r_2<2q\\
                      |s_2{}^q|\gcd(p,s_2{}^{r_2-2q}s)                & \phantom{2<\;}r_1=1 ~\land~ 2q< r_2\\
                      |p^{r_1-1}s_2{}^{r_2-q}|                        & 1\leq r_1\leq2 ~\land~ \phantom{2}q\leq r_2\leq 2q,
         \end{cases}\\
    k_3&=2|ps_2{}^{q+1}|k_2.
    \end{split}
    \end{equation}
    \end{widetext}
    In order to completely evaluate the remaining $\gcd$s we need to postulate
    \begin{equation}
    s_2=\pm p^u,\quad u\in\mathds{N}_0,
    \end{equation}
    so that
    \begin{widetext}
    \begin{equation}
    \begin{split}
    k_2&=\begin{cases}|p^{qu+1}|&\phantom{qu+1\leq\;}r_2u+r_1\leq2(qu+1)\\
                      |p^{(r_2u+r_1)-(qu+1)}|&qu+1\leq r_2u+r_1<2(qu+1),
         \end{cases}\\
    k_3&=2|p^{(q+1)u+1}|k_2.
    \end{split}
    \end{equation}
    \end{widetext}
\end{description}

\subsubsection*{\texorpdfstring{Case 2: $s_1\neq0\land ns_2=0$}{Case 2: s1 nonzero and ns2=0}}

We have
\begin{equation}
\begin{split}
k_0&=|s_1|,\\
k_1&=\sgn s_1,\\
k_2&=2\gcd(n,s_2Q_5)/|s_1|,\\
k_3&=2\gcd(n,s_2Q_5).
\end{split}
\end{equation}
In contrast to \autoref{sec:largeblacks} there are only three ways in which $ns_2$ can vanish.
\begin{description}
\item[Subcase 2.1: $s_2=0$.]
    \begin{equation}
    \begin{split}
    k_2&=2|n/s_1|,\\
    k_3&=2|n|.
    \end{split}
    \end{equation}
    Postulate
    \begin{gather}
    s_1=2^{p_1}n^{p_2}p_3,\quad p_{1,2}\in\mathds{N}_0,p_3\in\mathds{Z}\setminus\{0\},
    \shortintertext{under which}
    k_2=2^{1-p_1}|n^{1-p_2}p_3|,\quad p_1,p_2\in\{0,1\}.
    \end{gather}

\item[Subcase 2.2: $n=0$.]
    \begin{equation}
    \begin{split}
    k_2&=2|s_2Q_5/s_1|,\\
    k_3&=2|s_2Q_5|.
    \end{split}
    \end{equation}
    Postulate
    \begin{gather}
    \begin{gathered}
    l=2^{p_1}s_2{}^{p_2}Q_5{}^{p_3}p_4,\\
    p_{1,2,3}\in\mathds{N}_0,p_4\in\mathds{Z}\setminus\{0\},
    \end{gathered}
    \shortintertext{under which}
    \begin{gathered}
    k_2=2^{1-p_1}|s_2{}^{1-p_2}Q_5{}^{1-p_3}p_4|,\\
    p_1,p_2,p_3\in\{0,1\}.
    \end{gathered}
    \end{gather}

\item[Subcase 2.3: $n=s_2=0$.]
    \begin{equation}
    \begin{split}
    k_2&=1,\\
    k_3&=|s_1|.
    \end{split}
    \end{equation}
\end{description}

\subsubsection*{\texorpdfstring{Case 3: $ns_1s_2\neq0$}{Case 3: ns1ns2 nonzero}}

This time we are required to find solutions of $|s_1{}^2-4ns_2|= p^2,p\neq0$. As before, we first examine odd and even $s_1$:
\begin{description}
\item[Subcase 3.1: $s_1=2p$.] To force $k_0\in\mathds{Z}$ the product $ns_2$ needs to be a sum or difference of squares $p^2\pm q^2$, in which case
    \begin{equation}
    \begin{split}
    k_0&=2|q|\\
    k_1&=p/|q|.
    \end{split}
    \end{equation}
    \begin{description}
    \item[3.1.1] One way to achieve the desired form is through
        \begin{gather}
        s_2=p+q,\quad n=p-q,
        \shortintertext{under which}
        k_2=|q|^{-1}\gcd(p-q,(p+q)Q_5).
        \end{gather}
        Restricting to $q=\pm1$ to satisfy integer $k_1$ sets $k_0=2$ and $k_1=p$. The discrete invariant $d_1$ reduces to
        \begin{equation}
        \begin{split}
        d_1(\tilde{x})&=\gcd(p,\gcd(p\mp1,(p\pm1)Q_5))\\
                      &=\gcd(p,p\mp1,(p\pm1)Q_5)\\
                      &=1.
        \end{split}
        \end{equation}

    \item[3.1.2] Alternatively one may postulate
        \begin{gather}
        s_2=p^2\pm q^2,\quad n=1,
        \shortintertext{so that}
        \begin{split}
        k_2&=|q|^{-1}\gcd(1,(p^2\pm ^2q)Q_5)\\
        &=1/|q|,
        \end{split}
        \end{gather}
        and we must choose $q=\pm1$.

    \item[3.1.3] If one instead picks
        \begin{gather}
        s_2=1,\quad n=p^2\pm q^2,
        \shortintertext{we have}
        k_2=|q|^{-1}\gcd(p^2\pm ^2q,Q_5).
        \end{gather}
        This time it is not necessary to set $q=\pm1$ since $p$ and $Q_5$ may be multiples of $q$:
        \begin{gather}
        p=r_1q,\quad Q_5=r_2q,
        \shortintertext{so that}
        \begin{split}
        k_2&=\gcd((r_1\pm1)q,r_2)\\
        d_1(\tilde{x})&=\gcd(r_1,(r_1\pm1)q,r_2)\\
                      &=\gcd(q,r_1,r_2).
        \end{split}
        \end{gather}
        However if we do set $q=\pm1$ we obtain
        \begin{equation}
        \begin{split}
        k_2&=\gcd(p^2\pm1,Q_5)\\
        d_1(\tilde{x})&=\gcd(p,p\pm1,Q_5)\\
                      &=\gcd(p,1,Q_5)\\
                      &=1.
        \end{split}
        \end{equation}
    \end{description}

\item[Subcase 3.2: $s_1=2p+1$.] To counter the linear term arising in $s_1^2$ we propose
    \begin{gather}
    s_2=p,\quad n=p+1
    \shortintertext{under which}
    \begin{split}
    k_0&=1\\
    k_1&=2p+1\\
    k_2&=2\gcd(p+1,pQ_5).
    \end{split}
    \end{gather}
    One could just as well have chosen $s_2=p+1, n=p$, but this is trivially related to the chosen ansatz. Generalisations involving a second parameter $q$ run into the same $k_1$ obstacle as in subcase 3.1. We also find
    \begin{equation}
    \begin{split}
    d_1(\tilde{x})&=\gcd(2p+1,2\gcd(p+1,pQ_5))\\
                  &=\gcd(2p+1,2(p+1),2pQ_5)\\
                  &=1.
    \end{split}
    \end{equation}
\end{description}

Outside of these specialised cases one needs to solve the Diophantine equation $|l^2-4mn|=p^2,p\neq0$.

\section{Undoing an F-Duality with a U-Duality in the projective case}
\label{sec:FDualUDualUndo}

We know from \autoref{sec:gcds} and \cite{Krutelevich:2004} that all projective black holes of the same entropy are U-dual to each other (since U-duality acts transitively on the orbits). We also know that F-duality preserves entropy, hence when we consider the F-dual of a projective black hole, we must be able to ``U-dual it back'' to the original black hole. We show that this is true here.

Furthermore, as we saw in \autoref{sec:PQBasis}, under F-duality, we have
\begin{equation}
\begin{aligned}
P^2      &\to \sgn(\Delta) P^2 \\
Q^2      &\to \sgn(\Delta) Q^2 \\
P\cdot Q &\to \sgn(\Delta) P\cdot Q,
\end{aligned}
\end{equation}
but since $P$ and $Q$ transform under the $S\times T$ duality group $SL(2,\mathds{Z})\times SO(6,6;\mathds{Z})$, and an $S$ or a $T$ duality cannot flip the signs of $P^2$ or $Q^2$, the $U$-duality that undoes the F-duality \emph{must} be in the larger $E_{7(7)}$.

A general projective black hole has the form (see \eqref{eq:projective_fts})
\begin{equation}
x = \begin{pmatrix} 1 & (1,1,m) \\ (0,0,0) & j  \end{pmatrix},
\end{equation}
where $j\in\{0, 1\}$ and $m\in\mathds{Z}$. But the only examples that have a well defined Freudenthal dual have $m = \pm1$, $j = 0$ and $m = 0$, $j = 1$. Let us look at the first case.

We are to show that Freudenthal dual of
\begin{equation}
x = \begin{pmatrix} 1 & (1,1,m) \\ (0,0,0) & 0 \end{pmatrix},
\end{equation}
with $P^2= 2m$, $Q^2= -2$ and $P\cdot Q = 0$ given by
\begin{align}
\tilde{x} = \begin{pmatrix} 0 & (0,0,0) \\ (m,m,1) & m \end{pmatrix},
\end{align}
with $P^2= -2$, $Q^2= 2m$ and $P\cdot Q = 0$, is $U$-dual to $x$. We will use the U-dual transformations defined in \eqref{eq:FreudenthalConstructionTransformations}.

First we put the $B$ component of $\tilde{x}$ into Smith normal form\footnote{In general the operations used to put the Jordan algebra elements in Smith normal form do not lie in the U-duality group of the $STU$ model. However, in this particular example, they actually correspond to a triality.},
\begin{equation}
\begin{split}
\tilde{x} &=  \begin{pmatrix} 0 & (0,0,0) \\ (m,m,1) & m \end{pmatrix}  \\
          &\to\begin{pmatrix} 0 & (0,0,0) \\ (1,m,m) & m \end{pmatrix}.
\end{split}
\end{equation}
On the $P$s and $Q$s, this looks like
\begin{equation}
\begin{array}{c@{\ =\ }ccc@{\ =\ }c}
P^2      & -2 &                               & P^2      & -2m\\
Q^2      & 2m & \xrightarrow{\text{triality}} & Q^2      & 2  \\
P\cdot Q & 0  &                               & P\cdot Q & 0.
\end{array}
\end{equation}
So already, we see that for $m = -1$, a simple triality will flip the signs of $P^2$ and $Q^2$. Now apply a $\phi(C)$ transformation with $C = (1,0,0)$, then
\begin{gather}
\begin{split}
                     &\begin{pmatrix} 0 & (0,0,0) \\ (1,m,m) & m \end{pmatrix} \\
\xrightarrow{\phi(C)}&\begin{pmatrix} 1 & (m,0,0) \\ (1,m,m) & m \end{pmatrix}.
\end{split}
\shortintertext{for which}
\begin{array}{c@{\ =\ }ccc@{\ =\ }c}
P^2      & -2m &                       & P^2      & -2m  \\
Q^2      & 2   & \xrightarrow{\phi(C)} & Q^2      & 2-2m \\
P\cdot Q & 0   &                       & P\cdot Q & -2m.
\end{array}
\end{gather}
Followed by a $\psi(D)$ transformation with $D = (-1,-m,-m)$
\begin{equation}
\begin{split}
                     &\begin{pmatrix} 1 & (m,0,0)      \\ (1,m,m) & m    \end{pmatrix}\\
\xrightarrow{\psi(D)}&\begin{pmatrix} 1 & m(1-m,-1,-1) \\ (0,0,0) & 2m^2 \end{pmatrix}.
\end{split}
\end{equation}
Recall that $m=\pm1$ so $m^2=1$, so the last element looks like (after some triality)
\begin{gather}
\begin{pmatrix} 1 & (-m,-m,m-1) \\ (0,0,0) & 2 \end{pmatrix},
\shortintertext{for which}
\begin{array}{c@{\ =\ }ccc@{\ =\ }c}
P^2      & -2m  &                       & P^2      & m-2 \\
Q^2      & 2-2m & \xrightarrow{\psi(D)} & Q^2      & 2m  \\
P\cdot Q & -2m  &                       & P\cdot Q & -2.
\end{array}
\end{gather}
Now apply transformation (iii) from Lemma 27 of \cite{Krutelevich:2004} with $c=1$, to get
\begin{gather}
\begin{split}
                             &\begin{pmatrix} 1 & (-m,-m,m-1) \\ (0,0,0) & 2 \end{pmatrix}\\
\xrightarrow{\text{Lemma 27}}&\begin{pmatrix} 1 & (-m,-m,m)   \\ (0,0,0) & 0 \end{pmatrix},
\end{split}
\shortintertext{for which}
\begin{array}{c@{\ =\ }ccc@{\ =\ }c}
P^2      & m-2 &                               & P^2      & -2 \\
Q^2      & 2m  & \xrightarrow{\text{Lemma 27}} & Q^2      & 2m \\
P\cdot Q & -2  &                               & P\cdot Q & 0.
\end{array}
\end{gather}
For $m=-1$ then this is equal to the original $x$, for $m=+1$ we  have
\begin{equation}
\begin{pmatrix} 1 & (-1,-1,1) \\ (0,0,0) & 0 \end{pmatrix},
\end{equation}
to which we apply a norm preserving $T$ transformation such that the $A$ component goes to $(1,1,1)$, and then we are back to the original $x$,
\begin{gather}
\begin{pmatrix} 1 & (-1,-1,1) \\ (0,0,0) & 2 \end{pmatrix}\xrightarrow{T}
\begin{pmatrix} 1 & (1,1,1)   \\ (0,0,0) & 0 \end{pmatrix}
\shortintertext{with}
\begin{array}{c@{\ =\ }ccc@{\ =\ }c}
P^2      & -2 &                 & P^2      & 2m \\
Q^2      & 2m & \xrightarrow{T} & Q^2      & -2 \\
P\cdot Q & 0  &                 & P\cdot Q & 0.
\end{array}
\end{gather}

\section{Alternative Jordan Dual Formulation}
\label{sec:altJordanDual}

Recall from \autoref{sec:jdual} that we defined the Jordan dual $A^\star$ of $A$ as
\begin{equation}
A \to A^\star = \frac{A^\sharp}{N(A)^{1/3}},
\end{equation}
part of the motivation for this definition is that the entropy is preserved under J-Duality:
\begin{equation}
\begin{split}
N(A^\star) &= N\left(\frac{A^\sharp}{N(A)^{1/3}}\right) = \frac{1}{N(A)}N(A^\sharp)\\
           &= \frac{N(A)^2}{N(A)} = N(A).
\end{split}
\end{equation}

However, we note that, while $A$ belongs to the fundamental representation eg $\rep{27}$ of $E_6$ and describes a black string, $A^\star$ belongs to the contragredient representation eg $\rep{27'}$ of $E_6$ and corresponds to a black hole (the $^\sharp$ map is a map between the two representations).

An alternative definition which maps  $\rep{27}$  to  $\rep{27}$ and $\rep{27'}$  to  $\rep{27'}$ begins with a black string/hole pair. To lowest order, the extremal non-rotating black string and black hole entropies are given respectively by
\begin{equation}
S_5=(\pi\sqrt{|N(A)|},\pi\sqrt{|N(B)|}),
\end{equation}
where $N(A)= N(A,A,A)$. Large BPS and small BPS correspond to
$N(A) \neq 0$, and $N(A)=0$, respectively. The Dirac-Schwinger quantisation condition relating a black string/hole pair with charges $(A,B)$ to one with charges $(A',B')$ in the Jordan language is given by
\begin{equation}
\Tr(A, B')-\Tr(B,A') \in \mathds{Z}.
\end{equation}

The alternative \emph{Jordan dual} or J-dual, defined for ``large'' black strings and holes by
\begin{gather}
({A}^\star,  {B}^\star)=\pm \left (\frac{B^\sharp}{N(B)^{1/3}}, \frac{A^\sharp}{N(A)^{1/3}}\right),
\shortintertext{for which}
({A}^{\star\star},  {B}^{\star\star})=(A,B).
\end{gather}

In the case of a black string and a black hole related by J-duality
\begin{equation}
\Tr( B^\star, A)-\Tr(A^\star, B)=3(N(A)^{2/3}-N(B)^{2/3}).
\end{equation}
Note the factor of three. Hence, for a valid dual $({A}^\star,{B}^\star)$ we require that $N(A)^2$ and $N(B)^2$ are perfect cubes.  This is a necessary, but not sufficient condition because we also require that  ${A}^\star$ and ${B}^\star$ are themselves integer. This restricts us to that subset of black strings and holes for which
\begin{align}
d_3(\tilde{B})&=\left[ \frac{d_2(B)}{d_1({A}^\star)}\right]^3,&
d_3(\tilde{A})&=\left[ \frac{d_2(A)}{d_1({B}^\star)}\right]^3,
\end{align}
where $d_1(A)=\gcd (A)$, $d_2(A)= \gcd (A^\sharp)$ and $d_3(A)=|N(A)|$. Then
\begin{equation}
\begin{split}
 &\Tr( B^\star, A)-\Tr(A^\star, B)\\
=\ &3\left\{\left[ \frac{d_2(B)}{d_1({A}^\star)}\right]^2-\left[ \frac{d_2(A)}{d_1({B}^\star)}\right]^2\right\}.
\end{split}
\end{equation}
The U-duality integral invariants $\Tr(A,B)$ and $N(A,B,C)$ are not generally invariant under Jordan duality but $\Tr(B^\star, A)-\Tr(A^\star, B)$,  $N(A)$ and $N(B)$ and hence the lowest-order black string and black hole entropy, are invariant under this alternative J-duality but only up to an A-B interchange:
\begin{equation}
(N(A^\star),N(B^\star))=(N(B),N(A)).
\end{equation}

\providecommand{\href}[2]{#2}\begingroup\raggedright\endgroup

\end{document}